\documentclass[sigconf,natbib=true,screen=true]{acmart}

\usepackage{amsmath}
\usepackage{multirow}
\usepackage{graphicx}
\usepackage[inline]{enumitem}
\usepackage{sistyle}
\usepackage{algorithm}
\usepackage{algorithmic}
\usepackage{booktabs}
\usepackage{caption}
\SIthousandsep{,}
\usepackage{makecell}
\usepackage[skip=0pt]{caption}
\usepackage{algorithm}
\usepackage{algorithmic}
\usepackage{subfigure}

\AtBeginDocument{%
  \providecommand\BibTeX{{%
    \normalfont B\kern-0.5em{\scshape i\kern-0.25em b}\kern-0.8em\TeX}}}

\newcommand{\header}[1]{\vspace{1mm}\noindent\textbf{#1.}}

\copyrightyear{2022}
\acmYear{2022}
\setcopyright{acmcopyright}\acmConference[CIKM '22]{Proceedings of the 31st ACM International Conference on Information and Knowledge Management}{October 17--21, 2022}{Atlanta, GA, USA}
\acmBooktitle{Proceedings of the 31st ACM International Conference on Information and Knowledge Management (CIKM '22), October 17--21, 2022, Atlanta, GA, USA}
\acmPrice{15.00}
\acmDOI{10.1145/3511808.3557256}
\acmISBN{978-1-4503-9236-5/22/10}

\begin{document}

% alternative way of organizing the authors, from https://ftp.snt.utwente.nl/pub/software/tex/macros/latex/contrib/acmart/acmart.pdf, page 11
% \if 0
\author{Chen Wu}
\email{wuchen17z@ict.ac.cn}
\affiliation{
 \institution{CAS Key Lab of Network Data Science and Technology, ICT, CAS}
 \institution{University of Chinese Academy of Sciences}
 \city{Beijing}
 \country{China}
}

\author{Ruqing Zhang}
\email{zhangruqing@ict.ac.cn}
\author{Jiafeng Guo}
\authornote{Jiafeng Guo is the corresponding author.}
\email{guojiafeng@ict.ac.cn}
\affiliation{
 \institution{CAS Key Lab of Network Data Science and Technology, ICT, CAS}
 \institution{University of Chinese Academy of Sciences}
 \city{Beijing}
 \country{China}
}

\author{Wei Chen}
\email{chenwei2022@ict.ac.cn}
\author{Yixing Fan}
\email{fanyixing@ict.ac.cn}
\affiliation{
 \institution{CAS Key Lab of Network Data Science and Technology, ICT, CAS}
 \institution{University of Chinese Academy of Sciences}
 \city{Beijing}
 \country{China}
}

\author{Maarten de Rijke}
\orcid{0000-0002-1086-0202}
\email{m.derijke@uva.nl}
\affiliation{%
  \institution{University of Amsterdam}
  \city{Amsterdam}
  \country{The Netherlands}
}

\author{Xueqi Cheng}
\email{cxq@ict.ac.cn}
\affiliation{
 \institution{CAS Key Lab of Network Data Science and Technology, ICT, CAS}
 \institution{University of Chinese Academy of Sciences}
 \city{Beijing}
 \country{China}
}

\renewcommand{\shortauthors}{Chen Wu et al.}

\title{Certified Robustness to Word Substitution Ranking Attack for Neural Ranking Models}

\begin{abstract}
Neural ranking models (NRMs) have achieved promising results in  information retrieval. 
NRMs have also been shown to be vulnerable to adversarial examples. 
A typical Word Substitution Ranking Attack (WSRA) against NRMs was proposed recently, in which an attacker promotes a target document in rankings by adding human-imperceptible perturbations to its text. 
This raises concerns when deploying NRMs in real-world applications.  
Therefore, it is important to develop techniques that defend against such attacks for NRMs. 
In empirical defenses adversarial examples are found during training and used to augment the training set. 
However, such methods offer no theoretical guarantee on the models' robustness and may eventually be broken by other sophisticated WSRAs. 
To escape this arms race, rigorous and provable certified defense methods for NRMs are needed.

To this end, we first define the \textit{Certified Top-$K$ Robustness} for ranking models since users mainly care about the top ranked results in real-world scenarios. 
A ranking model is said to be Certified Top-$K$ Robust on a ranked list when it is guaranteed to keep documents that are out of the top $K$ away from the top $K$ under any attack. 
Then, we introduce a Certified Defense method, named CertDR, to achieve certified top-$K$ robustness against WSRA, based on the idea of randomized smoothing. 
Specifically, we first construct a smoothed ranker by applying random word substitutions on the documents, and then leverage the ranking property jointly with the statistical property of the ensemble to provably certify top-$K$ robustness. 
Extensive experiments on two representative web search datasets demonstrate that CertDR can significantly outperform  state-of-the-art empirical defense methods for ranking models. 
\end{abstract}




\maketitle

\section{Introduction}

Neural ranking models (NRMs) \cite{onal-neural-2018,Duet,dai2019deeper}, especially pre-trained ranking models \cite{monoBERT,ColBERT,ma2021b}, have led to substantial performance improvements in a wide range of search tasks \cite{ma2021b,joshi2020spanbert,gu2020speaker}. 
We have also seen NRMs being used in various practical usages in the enterprise \cite{lin2021pretrained}. 
Despite their success, recent observations \cite{wu2021neural, wu2022prada} show that NRMs are vulnerable to adversarial examples.
A typical word substitution ranking attack (WSRA) \cite{wu2022prada} was proposed and proved successful for attacking NRMs.
In this setting, an attacker promotes a target document in rankings by replacing important words in its text with their synonyms in a semantic-preserving way. 
The adversarial documents generated by WSRAs are imperceptible to humans but can easily fool NRMs. 
This may bring great concerns when deploying these models to real-world applications. 
Thus, efficient methods for defending against such attacks are of critical importance for deploying modern NRMs to practical search engines. 

Various approaches have been developed to defend against adversarial attacks. 
One representative type of defense is the empirical defense method \cite{madry2018towards,wang2019convergence}, where the set of perturbations is known at training time and adversarial examples are added to the training set \cite{wang2019convergence}. 
However, it is insufficient when considering all possible WSRAs in which each word in a document can be replaced with any of its synonyms. 
Consequently, such defenses may eventually be broken by other sophisticated WSRAs \cite{ye2020safer}.  
To escape this arms race,  procedures with rigorous and provable robustness guarantees are of special importance to the study of robustness of NRMs. 
In general, a model is said to be \emph{certified robust} if such an attack is guaranteed to fail, no matter how the attacker manipulates the input. 
A line of work on certified defenses against any admissible adversarial attack has recently been introduced in image and text classification \cite{cohen2019certified,katz2017reluplex,singh2019beyond}.  
However, existing certified defense approaches are mainly for  simple classification scenarios. 
They are quite distant from IR requirements, not just due to  input differences (discrete text vs.\ continuous image \cite{cohen2019certified}), but also due to the prediction type (ranked list vs.\ class label \cite{ye2020safer}). 
%Existing certified defenses usually consist of two steps, i.e., robustness verification to provide the theoretically certified lower bound of robustness under certain conditions, and corresponding robust training to improve such lower bound. 
In this sense, the recent, promising advances in certified defenses have yet to bring similar robustness guarantees in how NRMs are approached. 

Therefore, in this work, we make the first attempt to develop a certified defense method for NRMs so as to pursue adversarial immunization to WSRA to some extent.  
To this end, we need to  answer two key questions: 
First, what is certified robustness in IR? 
Second, can we train NRMs that are robust in this sense? 

\begin{figure}[t]
\centering
\includegraphics[width=\columnwidth]{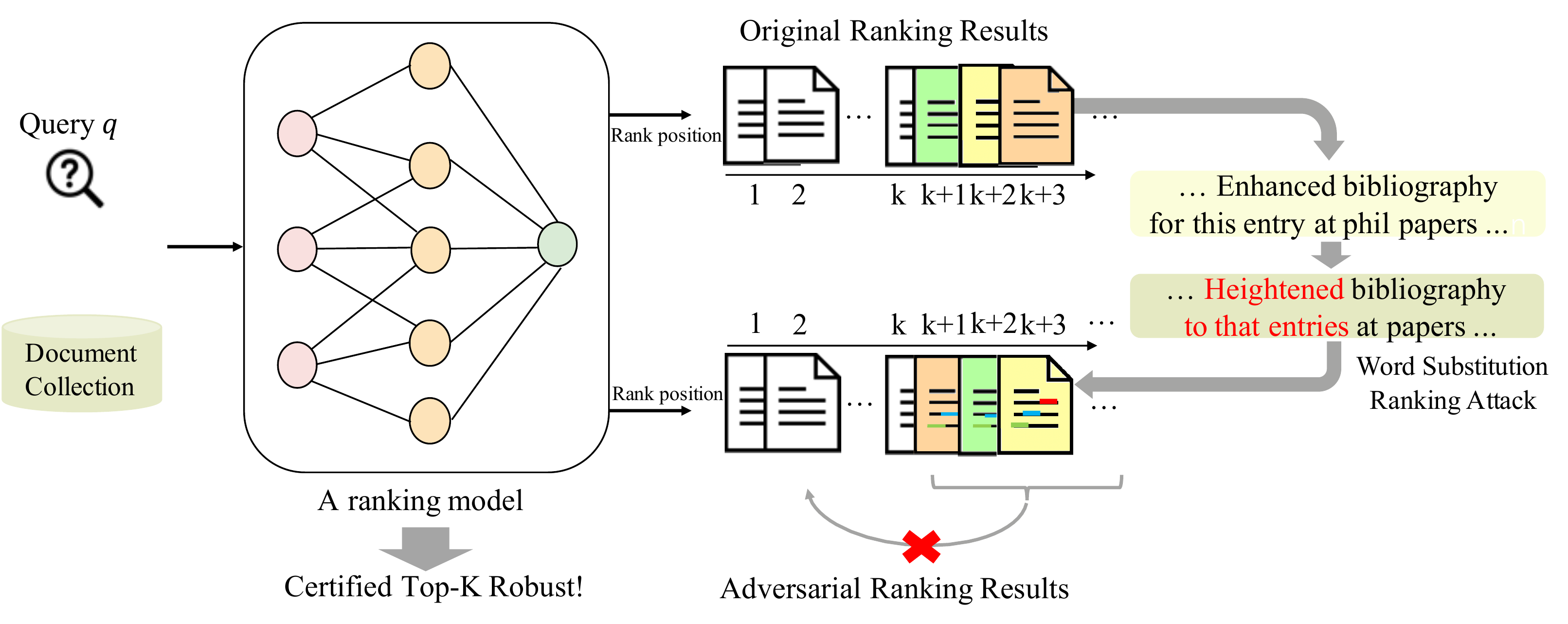}
\caption{Illustration of the WSRA and \textit{Certified Top-$K$ Robustness}. Given a ranking model, it is said to be Certified Top-$K$ Robust against WSRA on a ranked list if it is able to prevent documents outside top $K$ from appearing in top $K$ for all the possible WSRAs.}
\label{fig:certified top-k robustness}
\end{figure}

To answer the first question, based on the previous definition and inspired by the IR properties, we propose to define the \textit{Certified Top-$K$ Robustness} of ranking models. 
In IR, it seems well-accepted that in many scenarios users mainly care about top ranked results \cite{xia2009statistical, niu2012top}. 
Some observations have shown that traffic (or click-through rate) falls off steeply as users work their way down the search results \cite{joachims2017accurately}. 
Many widely-used ranking metrics (e.g., MRR and nDCG) are concentrated on top ranked predictions. 
Therefore, as illustrated in Figure \ref{fig:certified top-k robustness},
a ranking model is said to be \emph{Certified Top-$K$ Robust against WSRA} on a ranked list if it is able to prevent documents outside the top $K$ from appearing in top $K$ for all possible WSRAs.

To answer the second question, we propose a novel \textbf{Cert}ified \textbf{D}efense method for \textbf{R}anking models, CertDR for short, to enhance a model's certified robustness against WSRA. 
To avoid exponential computational costs, our method is based on the idea of randomized smoothing \cite{cohen2019certified, ye2020safer}, which replaces the ranking model with a smoothed ranker for which it is easier to verify the certified robustness. Specifically, we first construct a smoothed ranker by averaging the output ranking scores of randomly perturbed documents.  
Then, we obtain a certification criterion to judge models' certified top-$K$ robustness by leveraging the ranking property and statistical property of randomized ensembles. 
Finally, we design a practical certified defense algorithm, including a noise data augmentation strategy based on the perturbed documents for training and a rigorous statistical procedure to certify the top-$K$ robustness.

We conduct experiments on two web search benchmark datasets, i.e., the MS MARCO document ranking dataset and the MS MARCO passage ranking dataset. 
We first compare the certified robustness among different ranking models (i.e., traditional probabilistic ranking models, and advanced neural ranking models) under CertDR. 
Based on the evaluation results, there clearly remains room for future certified robustness improvements. 
Besides, we compare with several state-of-the-art empirical defense methods and our experimental results show that CertDR can achieve the best defense performance against WSRA.

\section{Related Work}
%In this section, we briefly review three lines of related work, text ranking models, adversarial attacks and defensive methods.

\subsection{Text Ranking Models}
%Information retrieval is a core task in many real-world applications, such as Web search and legal search.
Ranking models lie at the heart of IR.
Many different ranking models have been proposed over the past decades, including probabilistic models \cite{QL, BM25} (e.g., BM25 \cite{BM25}) and learning-to-rank (LTR) models \cite{liu2011learning}.
With the advance of deep learning, we have witnessed a substantial growth of interest in NRMs \cite{onal-neural-2018,Duet,dai2019deeper}, achieving promising results in a variety of search tasks \cite{ma2021b,joshi2020spanbert,gu2020speaker}.
Recently, pre-trained ranking models such as BERT-based models \cite{monoBERT,ColBERT} have shown substantial performance improvements both in academic research and industry \cite{lin2021pretrained}.
However, recent observations \cite{wu2021neural, wu2022prada} have shown that NRMs are vulnerable to adversarial examples.
In this paper, we study how to defend against adversarial attacks for NRMs.

\subsection{Adversarial Attacks}
Adversarial attacks aim to generate human-imperceptible adversarial examples by perturbing inputs to maximally increase a model's risk of making wrong predictions.
Adversarial examples were first discovered in the image domain \cite{szegedy2013intriguing}, where early research has developed powerful white-box attack methods, e.g., Fast Gradient Sign Method \citep[FGSM,][]{FGSM} and Projected Gradient Descent \citep[PGD,][]{PGD}, for attacking continuous image data. 

The existence and pervasiveness of adversarial examples have also been observed in the text domain \cite{zhang2020adversarial}.
Despite the fact that generating adversarial examples for texts has proven to be more challenging than for images due to their discrete nature, prior work has explored adversarial attacks for many language tasks, including text classification \cite{gao2018black,liang2017deep}, dialogue systems  \cite{cheng2019evaluating}, and natural language inference \cite{alzantot2018generating}. 
Among these attacks, word substitution attacks  \cite{alzantot2018generating,ren2019generating,zang2019word}, which replace words in a sentence with their synonyms via a synonym table, have attracted considerable attention.
The reason is they can preserve the syntactic and semantics of the original input to the most considerable extent and are very hard to discern, even from a human's perspective \cite{dong2020towards}. 

Document authors compete for more favorable positions in  rankings~\cite{goren2018ranking}.
The behavior is often referred to as search engine optimization (SEO) \cite{gyongyi2005web}, which includes ``illegitimate'' black hat SEO \cite{gyongyi2005web} (e.g., web spam \cite{castillo2011adversarial}) and ``legitimate'' white hat SEO \cite{gyongyi2005web}. Early work proposed methods to detect black-hat SEO behavior~\cite{ntoulas2006detecting,piskorski2008exploring}. 
Later, \citet{goren2018ranking} proposed to attack LTR by replacing a passage in a document with other passages to promote its ranking. 
Recently, \citet{wu2022prada} proposed word sub\-sti\-tution ranking attacks (WSRAs) to promote the ranking of a document, which easily escapes the detection of traditional anti-spamming methods. 
In this paper, we focus on defending against WSRAs.

\subsection{Defense Methods}

To defend against adversarial attacks, many defense methods have been proposed to make models more robust.
These approaches can be classified into \emph{empirical} defenses and \emph{certified} defenses. 
Empirical defenses attempt to make models empirically robust to known adversarial attacks; this has been extensively explored in image \cite{madry2018towards,wang2019convergence} and text classification~\cite{ye2020safer, jia2019certified}.
Data augmentation \cite{jia2017adversarial,ribeiro2018semantically} is a representative empirical defense by augmenting adversarial examples with the original training data. 
Since empirical defenses are only effective for certain attacks rather than all attacks, a competition emerges between adversarial attacks and defense methods. 

To solve the attack-defense dilemma, researchers resort to  certified defenses to make models provably robust to certain kinds of adversarial perturbations. 
\citet{jia2019certified} and \citet{huang2019achieving} first proposed to certify the robustness to adversarial word substitutions by leveraging Interval Bound Propagation (IBP \cite{IBP}) in NLP. 
These IBP-based methods are limited to continuous word embeddings and are not applicable to  subword-level models like BERT. 
\citet{ye2020safer} recently adopted randomized smoothing to certify the robustness to word substitution attacks, by turning the original classifier into a smoothed classifier by adding noise to the input. 
The final class prediction of the smoothed classifier is decided by majority voting over the noised inputs. 
Randomized smoothing is the only certification method that scales up to large-scale neural networks like BERT \cite{cohen2019certified} and provides tight bounds on large datasets.
But existing certified defenses are limited to simple classification scenarios and NRMs are less well studied.
Therefore, in this work, we develop a certified defense method for NRMs based on randomized smoothing.

\vspace*{-1mm}
\section{Certified Top-$K$ Robustness in IR}

We first introduce the WSRA attack we consider in this paper. 
Then, we introduce the definition of our proposed notion of Certified Top-$K$ Robustness for ranking models to such attacks.

\label{sec: task}

\vspace*{-1mm}
\subsection{Word Substitution Ranking Attack}

\textbf{Attacks in Web Search.}
The web search eco-system is, perhaps, the largest-scale adversarial setting in which search methods operate \cite{goren2020ranking}. 
For many queries in the web retrieval setting there exists an on-going ranking competition, i.e., many web document authors manipulate their documents deliberately to promote them in rankings \cite{goren2018ranking}.
This practice is often referred to as search engine optimization (SEO) \cite{gyongyi2005web}. 
The consequences of SEO are that the quality of search results may rapidly decrease since many irrelevant documents are ranked higher than they deserve.

Very recently, a typical black-box word substitution ranking attack (WSRA) \cite{wu2022prada} was proposed to simulate such  real-world ranking competitions. 
Specifically, WSRA could successfully attack NRMs by generating human-imperceptible adversarial documents for rank promotion. 
The synonymous word substitution it employs could maximally maintain the naturalness and semantic similarity of the original document, making it easy for the generated adversarial documents to evade spam detection. 
Due to the popularity of NRMs and the challenges of defending against human-imperceptible perturbations, we focus on WSRA attacks and design a corresponding defense in this paper.

\header{Notation} 
In ad-hoc retrieval, given a query $q$ and a set of document candidates $\mathcal{D} = \{d_1, d_2, \ldots, d_N \}$ selected from a collection $\mathcal{C}$,
a ranking model $f$ aims to predict the relevance score $\{f(q,d_n) : n=1,2,\ldots,N\}$ between every pair of query $q$ and candidate document  for ranking the whole candidate set. 
For example,  $f$ outputs the ranked list $L = [d_N, d_{N-1},\ldots, d_1]$ if it determines $f(q,d_N) > f(q,d_{N-1}) \cdots > f(q,d_1)$.
In this paper, we assume the ranking score $f(q, d_n)$ is the probability of relevance from 0 to 1 \cite{dai2019deeper}, which can be easily achieved by adding a sigmoid operation on the output given by the ranking model.

In WSRA, an attacker replaces the important words in the document with their synonyms by maximizing the adversarial ranking loss to promote the target document in rankings. 
The number of important words is a hyper-parameter. 
Specifically, for any word $w$, we consider a pre-defined synonym set $S_w$ containing the synonyms of $w$ (including $w$ itself). Following \citet{ye2020safer}, we assume the synonymous relation is symmetric, that is, $w$ is in the synonym set of all its synonyms. 
The synonym set $S_w$ can be built based on GLOVE \cite{pennington2014glove}.

\header{Definition 3.1.
($\delta$-Word Substitution Ranking Attack)}
For an input document $d= \{w_1, w_2, \ldots, w_M \} \in \mathcal{D}$, a \textit{$\delta$-word substitution ranking attack} constructs an adversarial document $d' = ( w_1', w_2', \ldots, w_M')$ by perturbing at most $\delta \cdot M$ ($\delta \leq 1$) words in $d$ to any of their synonyms $w_m' \in S_{w_m}$. 
We denote the candidate set of adversarial documents as $S_d$, i.e.,
$$ S_d := \{d': \|d' - d \|_0 / \|d\| \leq \delta \}, $$
where $\left\|d' - d\right\|_0 := \sum_{m=1}^{M} \mathbb{I}\{w_m' \neq w_m \}$ is the Hamming distance, with $\mathbb{I}\{\cdot \}$ the indicator function.
$\left\|d\right\|$ denotes the number of words in the document $d$ and $w_m' \in S_{w_m}$. 
Ideally, all $d' \in S_d$ have the same semantic meaning as $d$ for human judges, but their ranks may be promoted by the ranking model $f$. 
The goal of the attacker is to find $d' \in S_d$ such that $f(q,d')>f(q,d) $. 
Note that we do not attack the documents ranked from 1 to $K$, since there is no need to attack user's top search results.

\vspace*{-1mm}
\subsection{Definition of Certified Top-$K$ Robustness}

\textbf{Certified Top-$K$ Robustness.}
In general, a model is said to be \emph{certified robust} if an attack is guaranteed to fail, no matter how the attacker manipulates the input \cite{ye2020safer}. 
In a real web search scenario, it is known that users usually care much more about the top ranking results than others \cite{niu2012top}. For example, the traffic and click-through rate (CTR) both fall off as users work their way down the search results in major search engines:
while the first and second search results may achieve 36.4$\%$ and 12.5$\%$ CTR, the 10th search result may achieve a CTR of only 2.2$\%$.\footnote{https://www.smartinsights.com/search-engine-optimisation-seo/seo-analytics/the-number-one-spot-how-to-use-the-new-search-curve-ctr-data/} 
Moreover, many widely-used ranking metrics \cite{MRR, nDCG}  focus on the top-$K$ ranking results, e.g., MRR@K and nDCG@K.

Therefore, protecting the results ranked at the top positions is of great importance \cite{xia2009statistical, niu2012top}, not only for real-world applications, but also for the robustness guarantee of widely-used ranking metrics. 
Inspired by this, we define the \textit{Certified Top-$K$ Robustness} of ranking models in IR, where a ranking model $f$ is said to be certified robust at the ranked list $L$ if it is guaranteed that the documents ranked after top $K$ will not be attacked to be ranked into top $K$ in $L$. 
Since we focus on the WSRAs in this work, based on this basic definition, we further define \textit{Certified Top-$K$ Robustness to WSRA}.

\header{Definition 3.2.
(Certified Top-$K$ Robustness to WSRA)} 
Formally, a ranking model $f$ is said to be \textit{Certified Top-$K$ Robust} against WSRAs on the ranked list $L_q$ with respect to a query $q$ if it can keep all the document $d \in L_{q}[K+1: ]$ away from the top-$K$ for all the possible $\delta$-word substitution ranking attacks (as defined in Definition 3.1), i.e., \begin{equation}
Rank_{L_q}(f(q, d')) > K, \text{for all}~d \in L_{q}[K+1 :]~\text{and any} ~d' \in S_d, 
\label{equ: basic problem}
\end{equation}
where $Rank_{L_q}(f(q, d'))$ denotes the rank position of the adversarial document $d'$ in $L_q$ given by the ranking model $f$.
It is highly challenging to judge if $f$ is certified robust since all the candidate adversarial documents in $S_d$ should be checked and the size of possible perturbations grows exponentially with $\delta$. 
Following existing work \cite{ye2020safer, jia2019certified}, we mainly consider the worst case when $\delta=1$, which is the most challenging case.

\vspace*{-1mm}
\section{Our Certified Defense Method}
Based on the definition of certified top-$K$ robustness to WSRA, we introduce a novel \textbf{Cert}ified \textbf{D}efense method for \textbf{R}anking models (CertDR) to enhance the certified robustness. 
We first introduce a randomized smoothing function for ranking and how to use it to certify the robustness theoretically.
Then, we propose a practical certified defense algorithm for ranking models.
Proofs are at the end of this section.

\begin{figure*}[h]
\centering %0.43
\includegraphics[scale=0.38]{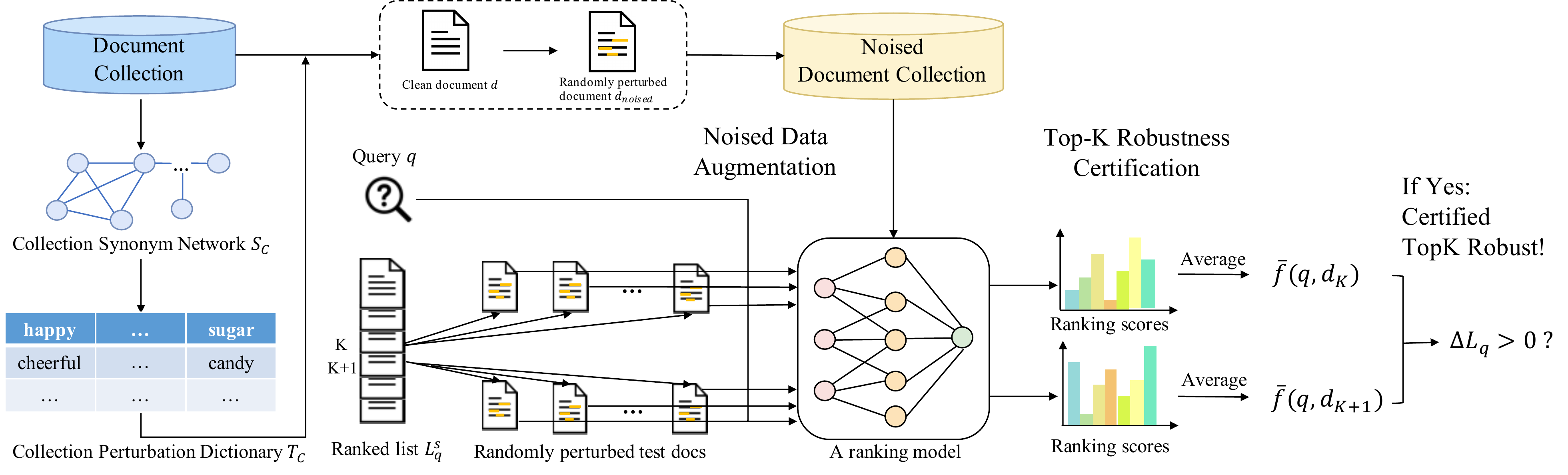}
% \vspace*{-2mm}
\caption{The overall architecture of the proposed practical certified defense method. 
We first generate the collection perturbation dictionary $T_C$ over the whole document collection.  
Then, we train the original ranking model by a noised data augmentation strategy to increase the robustness of $\bar{f}$.
Finally, we estimate the $\bar{f}(q,d_K)$ and $\bar{f}(q,d_{K+1})$ by Monte Carlo estimation.
The criterion $\Delta{L_q}$ is computed, where if $\Delta{L_q} > 0$, we can certify that $\bar{f}$ is top-$K$ robust at the ranked list $L_q^s$. }
\label{fig:overview}
\end{figure*}

\vspace*{-1mm}
\subsection{Randomized Smoothing Function for Ranking}

To circumvent the computationally expensive combinatorial optimization (e.g., enumerating all the candidate adversarial documents in $S_d$), we borrow the idea from the randomized smoothing technique \cite{cohen2019certified, ye2020safer}, 
which could provably defend against the adversarial attacks by leveraging the voting of randomly perturbed samples derived from the original input.
% which has been shown to scale up to large-scale neural networks and provide tight bound on large datasets. 
We target to replace the ranking model $f$ with a smoothed ranking model $\bar{f}$ for which it is easier to verify the certified robustness. 

Specifically, we construct the smoothed ranker $\bar{f}$ by averaging the output ranking scores of a set of randomly perturbed documents based on random perturbations, i.e.,
$$
\bar{f}(q, d) = \mathbb{E}_{\mathbf{R}\sim \Pi_d} f(q, \mathbf{R}),
$$
where $R$ is a randomly perturbed document and $\Pi_d$ is the corresponding probability distribution that prescribes a random perturbation around $d$. 
In our work, we define $\Pi_d$ to be the uniform distribution on a set of random word substitutions following \cite{ye2020safer}.

In previous classification tasks \cite{cohen2019certified, ye2020safer}, the output of the smoothed classifier is the class with the largest probability ``voting'' by all randomly perturbed inputs. 
Different from these works, we compute the output of the smoothed ranker by averaging the ranking scores of all randomly perturbed documents originated from the $d$, which is more suitable for the ranking problem. 
In this way, we could obtain the ranked list $L_q^s$ based on the averaged scores produced by the smoothed ranker $\bar{f}$. 
Here, we leave the query $q$ free from attack.  
In the future work, we would like to explore the defense against query attacks by focusing on $q$ in this formulation.

To obtain random perturbations in defense methods effectively, we propose to build a perturbation set $T_w$ for each word $w$. 
Specifically, we construct $T_w$ from the synonym set $S_w$ used in the attack method, i.e., WSRA in this work, by choosing the top $J$ nearest neighbors via the cosine similarity of GLOVE vectors. Then, for a document $d = (w_1, w_2, \ldots, w_M)$, we define the  perturbation distribution $\Pi_{d}$ by perturbing each word $w_i$ in $d$ to a word in its perturbation set $T_{w_i}$ randomly and independently, i.e., 
$$
\Pi_{d} (\mathbf{R}) = \prod_{i=1}^{M}\frac{\mathbb{I}\{r_i \in T_{w_i}\} }{|T_{w_i}|},
$$
where $\mathbf{R} = (r_1, \ldots, r_M)$ is the perturbed document and $|T_{w_i}|$ denotes the size of $T_{w_i}$. 

\vspace*{-1mm}
\subsection{Certifying Smoothed Ranking Models}
\label{sec: Certifying Smoothed Ranking Models}

Given the smoothed ranking model $\bar{f}$, in this section, we introduce how to certify its top-$K$ robustness. 
For all the documents in $L_q^s[K+1:]$, if their adversarial documents could achieve lower scores than the document $d_K$ ranked at the position $K$, we think these documents $L_q^s[K+1:]$ cannot be attacked into top K. 
Formally, the condition that $\bar{f}$ is certified top-$K$ robust on $L_q^s$ can be defined as, 
\begin{equation}
\max_{d\in L_q^s[K+1:]} \max_{d'\in S_d} \bar{f}(q,d') < \bar{f}(q, d_K). 
\label{equ:smoothed basic problem}
\end{equation}

\noindent%
In general, there are two difficulties to complete the above certification case by case, i.e., the inner maximum and the outer maximum. 

\header{Inner Maximum} The first difficulty is to exam all  candidate adversarial documents in $S_d$ for the inner maximum, where the computation cost grows exponentially with the attacked word number $\delta  M$. 
We address this problem in the following Theorem~\ref{thm:4.1}.  

\begin{theorem}[Certified Upper Bound] 
\label{thm:4.1}
Assume that the perturbation set $T_w$ is constructed such that $|T_w| = |T_{w'}|$ for every word $w$ and its synonym $w' \in S_w$. Define
$$ o_w = \min_{w' \in S_{w}} \frac{|T_w \cap T_{w'} |}{|T_w|},$$
where $o_w$ indicates the overlap between the two different perturbation sets.
For a given document $d = (w_1, \ldots, w_M)$, we sort the words according to $o_w$, such that $o_{w_{i_1}} \leq o_{w_{i_2}} \leq \cdots \leq o_{w_{i_M}}$. Then
% $$\min_{d' \in S_d}\bar{f}(q,d') \geq \max(\bar{f}(q,d) - o_d, 0)$$
$$\max_{d' \in S_d}\bar{f}(q,d') \leq \min(\bar{f}(q,d) + o_d, 1),$$
where $o_d := 1 - \prod_{j=1}^E o_{w_{i_j}}$. 
\end{theorem}

\noindent%
The idea is that for any adversarial document $d' \in S_d$, the upper bound of $\bar{f}(q,d')$ can be bounded by its original document ranking score $\bar{f}(q,d)$ with randomized smoothing.
As a result, this theorem avoids the difficult adversarial optimization of $\bar{f}(q, d') $ on $d' \in S_d$, and only needs to evaluate $\bar{f}(q,d)$ at the original document $d$. 
Note that the difference between our Theorem 4.1 with Theorem 1 in \cite{ye2020safer} is that we extend the classification prediction in Theorem 1 to ranking scores between queries and documents,
and prove the theorem in an upper bound situation, which has not been provided by \citet{ye2020safer}.
The proof is provided in Section \ref{Proof of Theorem 4.1}. 
Besides, the upper bound in Theorem~\ref{thm:4.1} is sufficiently tight, which is shown in Section~\ref{sec: Tightness}. 

\header{Outer Maximum} The second difficulty is to exam all documents in $L_q^s[K+1:]$ for the outer maximum, where the computational costs grow linearly with the list length $N$. 
We address the outer maximum in the following.
To simplify our notation, we define $A_L = \bar{f}(q, d_K) - \max_{d\in L_q^s[K+1:]} \max_{d'\in S_d} \bar{f}(q,d') $, where Eq.~(\ref{equ:smoothed basic problem}) is equivalent to $A_L > 0$. 
By applying Theorem~\ref{thm:4.1} to Eq.~(\ref{equ:smoothed basic problem}), we have
$$
\begin{aligned}
A_L &\geq \bar{f}(q, d_K) - \max_{d\in L_q^s[K+1:]} (\bar{f}(q,d) + o_d)   \\
&\geq \bar{f}(q, d_K) - \bar{f}(q,d_{K+1}) - \max_{d\in L_q^s[K+1:]}o_d ,
\end{aligned}
$$
where $d_{K+1}$ denotes the document which is ranked at the position $K+1$. 
The proof is achieved by utilizing the ranking property that $\bar{f}(q,d_1) > \bar{f}(q,d_2) > \cdot \cdot \cdot$.
The idea is that we can compute $A_L$ (in other words, certifying  $\bar{f}$) by comparing the ranking scores of documents ranked at $K$ and $K+1$. 
Note that the computational cost of $\max_{d\in L_q^s[K+1:]}o_d$ is negligible.
% much lower than $\max_{d\in L[K+1:]} (\bar{f}(q,d) + o_d)$ with $O(N)$ inferences.  

Based on the above solutions of the inner and outer maximum in  Eq. (\ref{equ:smoothed basic problem}), it is possible to introduce a certification criterion for checking the certified top-$K$ robustness for ranking models. 

\begin{proposition}
\label{prop:4.1}
For a ranked list $L_q^s$ with respect to a query $q$, under the condition of Theorem~\ref{thm:4.1}, we can certify that $Rank_{L_q^s}(\bar{f}(q, d')) > K, \text{for all}~d \in L_{q}^s[K+1 :]~\text{and any} ~d' \in S_d$ if 
\begin{equation}
\label{eq: cr criterion}
    \Delta L_q \overset{\text{def}}{=} \bar{f}(q, d_K) - \bar{f}(q,d_{K+1}) - \max_{d\in L_q^s[K+1:]}o_d > 0, 
\end{equation}
\end{proposition}
\noindent
where $\Delta L_q$ can be estimated by Monte Carlo estimation as we show in the next section. 
We can certify whether the ranking model is top-$K$ robust on the ranked list $L_q^s$ by simply checking $\Delta L_q$. 
If $\Delta L_q$ is positive, the model is certified top-$K$ robust. %Otherwise, it is not certified top-$K$ robust. 

\vspace*{-1mm}
\subsection{Practical Certified Defense Algorithm}

% key step : noise training + certify the robustness

Based on the above theoretical analysis, we now present a practical certified defense algorithm for ranking models. 
Formally, we write $S_C$ for the synonym dictionary of the collection.  
Specifically, $S_C$ contains the synonym set $S_w$ for all words in the collection $C$ and is often presented as a synonym network \cite{ye2020safer}. 
Similar to the process of obtaining $T_w$ from $S_w$, we achieve the collection perturbation dictionary $T_C$ by keeping the top $J$ nearest neighbors in $S_C$ for each word. The overall architecture is shown in Figure \ref{fig:overview}. 
%A pseudo algorithm is provided in Algorithm \ref{alg}. 

The certified defense algorithm contains two key steps, i.e., noised data augmentation and Top-$K$ Robustness Certification.  
We describe the two steps in the following.

\header{Noise Data Augmentation Strategy}
% 原因
The robustness certification holds regardless of how the original  ranker $f$ is trained. 
However, to rank the document $d$ with respect to the $q$ correctly and robustly by $\bar{f}$, it is expected that $f$  properly ranks the perturbed document $R$ (recall that $R\sim \Pi_d $) such that it is close to the rank position of the original document $d$.
That is, the noise of $R$ should have little effect on the ranking, making the ensembled ranking score $\bar{f}(q,d)$ close to the original ranking score $f(q,d)$.
However, if $f$ is trained via standard supervised learning without any noised documents, it will not necessarily learn how to rank $R$ properly. 

\iffalse
\begin{algorithm}[t]
 \caption{Practical Certified Defense Algorithm}
 \label{alg}
 \begin{algorithmic}[1] %1 
 \REQUIRE The original ranker $f$, the document collection $\mathcal{C}$, a ranked list $L_q$ with respect to the query $q$, the collection synonym dictionary $S_C$.
 \STATE Generate the collection perturbation dictionary $T_C$ from the whole document collection $C$.
 \STATE \textbf{Procedure} Noised Data Augmentation
 \FOR{$d \in \mathcal{C}$}
 \STATE Generate the randomly perturbed document $d_{noised}$.
 \ENDFOR
 \STATE Train a more robust ranking model starting from $f$ via  the noised documents, w.r.t. Eq.~(\ref{equ: noised training}), in order to obtain a better $\bar{f}$ with the ranked list $L_q^s$.
 \STATE \textbf{Procedure} Top-$K$ Robustness Certification
 \STATE Estimate $\bar{f}(q,d_K)$ and $\bar{f}(q,d_{K+1})$ by Monte Carlo estimation. 
 \STATE Compute $\Delta L_q$, w.r.t. Eq.~(\ref{eq: cr criterion})
 \IF{$\Delta L_q$ > 0}
 \STATE $\bar{f}$ is certified top-$K$ robust at the ranked list $L_q^s$.
 \ELSE
 \STATE $\bar{f}$ is not certified top-$K$ robust at the ranked list $L_q^s$.
 \ENDIF
 \end{algorithmic}
\end{algorithm}
\fi

% 做法
Inspired by previous works \cite{lecuyer2019certified, ye2020safer,zhou2020adversarial}, we introduce a noise data augmentation strategy for ranking. 
Specifically, we first generate a perturbed document $d_{noised}$ for each $d$ in the collection $C$. 
The perturbation is achieved by randomly sampling every word from $d$ using the perturbation distribution $\Pi_d$. 
Then, we train the original ranker $f$ using the training triples equipped with the noised documents via the following objective:
\begin{equation}
\label{equ: noised training}
L_{ndat} = \max (0, 1 - f(q, d_{noised}^{+}) + f(q, d_{noised}^{-})),
\end{equation}
where $d_{noised}^{+} / d_{noised}^{-}$ is the perturbed document from $d^{+} / d^{-}$. And $d^{+} / d^{-}$ denotes the positive/negative document in original training triples. 
Then, we can obtain a better smoothed ranker $\bar{f}$ by Monte Carlo estimation in the following.

\header{Top-$K$ Robustness Certification} 
In theory, since the perturbation space can be extremely large, it is impossible to exactly obtain the prediction of $\bar{f}$ at each $(q,d)$.
Therefore, based on the ranker $f$ obtained from the noised data augmentation strategy, we estimate $\bar{f}(q,d_K)$ and $\bar{f}(q,d_{K+1})$ by Monte Carlo estimation.
Take $\bar{f}(q,d_K)$ as an example, we can estimate it like
$$\bar{f}(q,d_K) = \mathbb{E}_{\mathbf{R}_K\sim d_K} f(q, d_K) \approx \frac{1}{n}\sum_{i=1}^{n}f(q,\mathbf{R}_K^{(i)}), $$
where $\mathbf{R}_K^{(i)}$ are i.i.d. samples from $\Pi_{d_K}$ and thus $\Delta L_q$ can be approximated accordingly.
We can construct rigorous statistical procedures to reject the null hypothesis that $\bar{f}$ is not certified robust at $L_q$ (e.g., $\Delta L_q < 0$) with a given significance level (e.g., 5\%) following \cite{ye2020safer}.

Finally, if $\Delta L_q >0$, then $\bar{f}$ is certified top-$K$ robust at the ranked list $L_q^s$. 
Otherwise, we will judge it is not certified top-$K$ robust at the $L_q^s$.
We can see that our practical certified defense algorithm could be achieved by assembling the ranking outputs and that it does not require any further information about the ranking models.
Thus, it can be applied to any ranking model.

\vspace*{-1mm}

\subsection{Proofs}
\label{sec: tightness and proofs}
Here we provide all the necessary proofs of Theorem~\ref{thm:4.1} and the tightness of the bound in Theorem~\ref{thm:4.1} with its proof. 
The upper bound of Theorem~\ref{thm:4.1} is achieved by introducing an auxiliary function cluster based on the relevance between the query and document, and solving the constraint optimization problem by Lagrange and  properties of randomly perturbed sets. 
Tightness is proved by constructing the randomized smoothing ranker that satisfies the desired property we want.

\vspace*{-2mm}
\subsubsection{Proof of Theorem 4.1}
\label{Proof of Theorem 4.1}
Our goal is to calculate the upper bound $\max_{d' \in S_d}\bar{f}(q,d')$. 
The key idea is to frame the computation of the upper bound into a variational optimization.

\begin{lemma}
\label{lem:4.1} Define $\mathcal{G}_{[0,1]}$ to be the set of all bounded functions mapping from $\mathcal{Q} \times \mathcal{D}$ to [0, 1].
For any $g \in \mathcal{G}_{[0,1]}$, define 
$$\Pi_d[g] = \mathbb{E}_{\mathbf{R}\sim \Pi_d}[g(q,\mathbf{R})].$$
Then we have for any $\mathbf{R}$, 
$$\begin{aligned}
\max_{d' \in S_d} \bar{f}(q, d') &\leq \max_{d' \in S_d} \max_{g \in \mathcal{G}_{[0,1]}} \{ \Pi_{d'}[g]~s.t.~\Pi_{d}[g]=\bar{f}(q,d) \} \\ &:=\bar{f}_{up}(q,d')
\end{aligned}
$$
\end{lemma}

\header{\rm\em Proof of Lemma~\ref{lem:4.1}} 
Define $g_{0}(q,d) = f(q,d)$. Then we have
$$\bar{f}(q,d)=\mathbb{E}_{\mathbf{R}\sim \Pi_d}[f(q,\mathbf{R})] = \Pi_d[g_{0}].$$
Therefore, $g_0$ satisfies the constraints in the optimization, which makes it obvious that 
$$\bar{f}(q,d') = \Pi_{d'}[g_0] \leq \max_{g \in \mathcal{G}_{[0, 1]}} \{\Pi_{d'}[g]~s.t.~\Pi_d[g]=\bar{f}(q,d) \}.$$
Taking $\max_{d' \in S_d}$ on both sides yields the upper bound and thus the problem reduces to deriving bounds for the optimization problems.

\begin{theorem}
\label{thm:4.3}
Under the assumption of Theorem~\ref{thm:4.1}, for the optimization problem in Lemma~\ref{lem:4.1}, we have
$$\bar{f}_{up}(q,d') \leq \min(\bar{f}(q,d) + o_d, 1), $$
where $o_d$ is the quantity defined in Theorem~\ref{thm:4.1}.
\end{theorem} 

\noindent%
%Now we proceed to prove Theorem~\ref{thm:4.3}.

\header{\rm\em Proof of Theorem~\ref{thm:4.3}}
For notation, we denote $p=\bar{f}(q,d)$.
Applying the Lagrange multiplier to the constraint optimization problem and exchanging the min and max, we have
$$
\begin{aligned}
\bar{f}_{up}(q,d') 
&= \max_{d' \in S_d} \max_{g \in \mathcal{G}_{[0,1]}} \{\Pi_{d'}[g]~s.t.~\Pi_d[g]=\bar{f}(q,d) \}   \\
&= - \min_{d' \in S_d} \min_{g \in \mathcal{G}_{[0,1]}}\{-\Pi_{d'}[g]~s.t.~\Pi_d[g]=\bar{f}(q,d) \}   \\
&\leq -\min_{d' \in S_d} \min_{g \in \mathcal{G}_{[0,1]}} \max_{\lambda \in \mathcal{R}}  \lambda \Pi_d[g] - \Pi_{d'}[g] - \lambda p  \\
&= \max_{d' \in S_d} \min_{\lambda \in \mathcal{R}} \max_{g \in \mathcal{G}_{[0,1]}} \Pi_{d'}[g] - \lambda \Pi_d[g] + \lambda p  \\ 
% &= \min_{\lambda \in \mathcal{R}} \max_{d' \in S_d}  \max_{g \in \mathcal{G}_{[0,1]}} \int g(\mathbf{R})(d\Pi_{d'}(\mathbf{R}) - \lambda d \Pi_d(\mathbf{R})) + \lambda p  \\
&= \min_{\lambda \in \mathcal{R}} \lambda p + \max_{d' \in S_d}  \max_{g \in \mathcal{G}_{[0,1]}} \int g(\mathbf{R})(d\Pi_{d'}(\mathbf{R}) - \lambda d \Pi_d(\mathbf{R})).  
\end{aligned}
$$

\noindent%
Note that 
$$\max_{g \in \mathcal{G}_{[0,1]}} \int g(\mathbf{R})(d\Pi_{d'}(\mathbf{R}) - \lambda d \Pi_d(\mathbf{R})) =  \int (d\Pi_{d'}(\mathbf{R}) - \lambda d \Pi_d(\mathbf{R}))_{+}, $$
which is achieved by setting
$$
g(\mathbf{R})
 = \mathbb{I}\{d\Pi_{d'}(\mathbf{R}) - \lambda d \Pi_d(\mathbf{R}) \geq 0 \},
$$
where $(a)_{+} = \max(a, 0)$ and $\mathbb{I}\{\cdot\}$ denotes the indicator function.
Thus we have,
$$
\begin{aligned}
\bar{f}_{up}(q,d') 
&= \min_{\lambda \in \mathcal{R}} \lambda p +  \max_{d' \in S_d}  \int (d\Pi_{d'}(\mathbf{R}) - \lambda d \Pi_d(\mathbf{R}))_{+}.
\end{aligned}
$$
For any $\lambda < 0$, we can show that
$$
\small
\begin{aligned}
\lambda p + \max_{d' \in S_d} \int (d\Pi_{d'}(\mathbf{R})& - \lambda d  \Pi_d(\mathbf{R}))_{+} \\
&= \lambda p + \max_{d' \in S_d} \int(d\Pi_{d'}(\mathbf{R}) - \lambda d \Pi_d(\mathbf{R}))\\
&= \lambda(p - 1) + 1 > 0(p-1) + 1 =1,
\end{aligned}
$$
which contradicts $\bar{f}(q,d') \leq 1$.
This implies that the minimum of $\lambda p +  \max_{d' \in S_d}  \int (d\Pi_{d'}(\mathbf{R}) - \lambda d \Pi_d(\mathbf{R}))_{+} $ must be achieved at $\lambda \geq 0$.
Thus we have
$$
\bar{f}_{up}(q,d') = \min_{\lambda \geq 0} \lambda p +  \max_{d' \in S_d}  \int (d\Pi_{d'}(\mathbf{R}) - \lambda d \Pi_d(\mathbf{R}))_{+}.
$$
Now we calculate $\int (d\Pi_{d'}(\mathbf{R}) - \lambda d \Pi_d(\mathbf{R}))_{+}$.

\begin{lemma}
\label{lem:4.2} 
Given the words $w$, $w'$, 
we write $n_w = |T_w|, n_{w'} = |T_w'|$, and $n_{w,w'} = |T_w \cap T_{w'}|$.
We have the following identify
$$
\small
\begin{aligned}
\int (d\Pi_{d'}(\mathbf{R}) &- \lambda d \Pi_d(\mathbf{R}))_{+} =  1 - \prod_{j\in [M], w_j \neq w'_j} \frac{n_{w_j,w'_j}}{n_{w'_j}}  \\
&+ \left(\prod_{j \in [M], w_j \neq w'_j} \frac{n_{w_j,w'_j}}{n_{w'_j}}\right) \left(1 - \lambda \prod_{j \in [M], w_j \neq w'_j} \frac{n_{w'_j}}{n_{w_j}} \right)_{+}.
\end{aligned}
$$
\end{lemma}

\noindent%
As a result, under the assumption that $n_w = |T_w| = |T_{w'}| = n_{w'}$ for every word $w$ and its synonym $w' \in S_w$, 
we have 
$$
\begin{aligned}
\int (d\Pi_{d'}(\mathbf{R}) - \lambda d \Pi_d(\mathbf{R}))_{+} &= 1 - \prod_{j\in [M], w_j \neq w'_j} \frac{n_{w_j,w'_j}}{n_{w'_j}} \\ &+ \left(\prod_{j \in [M], w_j \neq w'_j} \frac{n_{w_j,w'_j}}{n_{w'_j}}\right) \left(1 - \lambda \right)_{+}.
\end{aligned}
$$
Now we need to solve the optimization of  $\max_{d' \in S_d}  \int (d\Pi_{d'}(\mathbf{R}) - \lambda d \Pi_d(\mathbf{R}))_{+} $.

\begin{lemma}
\label{lem:4.3}
% 交集越小的会排在越前面， 这样就先取到小的ffffss
For any word $w$, define $\Tilde{w}^* = \arg \min_{w' \in S_d} n_{w,w'} / n_w $. For a given document $d = (w_1, \ldots, w_M)$, we define an ordering of the words $w_{p_1}, \ldots, w_{p_W}$ such that $n_{w_{p_i}, \Tilde{w}_{p_i}^*} / n_{w_{p_i}} \leq n_{w_{p_j}, \Tilde{w}_{p_j}^*} / n_{w_{p_j}}$ for any $i \leq j$. For a given $d$ and $E=\delta M$, we define an adversarial perturbed document $d^* = (w_1^*, \ldots, w_M^*)$, where 
$$ w_i^*=\left\{
                \begin{array}{ll}
                  \Tilde{w}^*, &\text{if } i \in [w_1, \ldots, w_E]\\
                  w_i, &\text{if } i \notin [w_1, \ldots, w_E]
                \end{array}
              \right. $$
Then for any $\lambda > 0$, we have that $d^*$ is the optimal solution of $\max_{d' \in S_d}  \int (d\Pi_{d'}(\mathbf{R}) - \lambda d \Pi_d(\mathbf{R}))_{+}$, that is,
$$
\max_{d' \in S_d}  \int (d\Pi_{d'}(\mathbf{R}) - \lambda d \Pi_d(\mathbf{R}))_{+} 
=   \int (d\Pi_{d^*}(\mathbf{R}) - \lambda d \Pi_d(\mathbf{R}))_{+}.
$$
\end{lemma}

\noindent%
Now by Lemma~\ref{lem:4.3}, the upper bound becomes
\begin{equation}
\begin{aligned}
\bar{f}_{up}(q,d') 
&= \min_{\lambda \geq 0} \lambda p +  \max_{d' \in S_d}  \int (d\Pi_{d'}(\mathbf{R}) - \lambda d \Pi_d(\mathbf{R}))_{+} \\
&= \min_{\lambda \geq 0} \lambda p +   \int (d\Pi_{d^*}(\mathbf{R}) - \lambda d \Pi_d(\mathbf{R}))_{+} \\
&= \min_{\lambda \geq 0} (\lambda p + o_d + (1-o_d)(1-\lambda)_{+}) \\
&= \min ( p + o_d, 1),
\end{aligned}
\label{equ:final upper bound}
\end{equation}
% 分类讨论的过程省略了，篇幅不够可以补上
where $o_d$ is consistent with the definition in Theorem~\ref{thm:4.1}:
$$
o_d = 1 - \prod_{j \in [M], w_j \neq \Tilde{w}^*_j}\frac{n_{w_j,\Tilde{w}^*_j}}{n_{w_j}} = 1 - \prod_{j=1}^{E}o_{w_{i_j}}.
$$

\noindent%
Here, Eq.~(\ref{equ:final upper bound}) is calculated using the assumption of Theorem~\ref{thm:4.1}. The optimization of $\min_{\lambda \geq 0}$ in (\ref{equ:final upper bound}) is an elementary step: if $p + o_d > 1$, we have $\lambda^* = 0$ with solution 1; if $p + o_d \leq 1$, we have $\lambda^* = 1$ with solution $p + o_d$. 
For the proof of Lemma~\ref{lem:4.2} and \ref{lem:4.3}, we 
 refer readers to \cite[][Lemma 2 and 3]{ye2020safer}.

\subsubsection{Tightness}
\label{sec: Tightness}
Whether the bound in Theorem~\ref{thm:4.1} is sufficiently tight is of great importance. 
In the following, we provide a theorem to state its tightness. 

\begin{theorem}[Tightness]
\label{thm:4.2}
Assume the conditions of Theorem~\ref{thm:4.1} hold. For a ranking model $f$ that maps $Q \times D$ to [0, 1], there exists a model $f_*$ such that its related smoothed ranker $\bar{f}_*$ satisfies 
$$
\bar{f}_*(q, d) = \bar{f}(q,d),
$$
and
$$
\max_{d' \in S_d} \bar{f}_*(q, d') = \min(\bar{f}_*(q,d) + o_d, 1),
$$
where $o_d$ is defined in Theorem~\ref{thm:4.1}.
\end{theorem}

\noindent%
As shown in Theorem \ref{thm:4.2}, the upper bound in Theorem~\ref{thm:4.1} is tight and can not be further improved if we do not know any other structural information about $f$. In the following, we provide the proof of Theorem \ref{thm:4.2}. 
% It is guaranteed by the randomized smoothing theory \cite{cohen2019certified} which can be achieved by only evaluating the output of $\bar{f}(q,d)$.

% if we access $\bar{f}$ only through the evaluation of $\bar{f}(q,d)$, then the bound in Theorem~\ref{thm:4.1} is the tightest possible that we can achieve, because we can not distinguish between $\bar{f}$ and $\bar{f}_*$ in Theorem~\ref{thm:4.2} with the information available. 

\header{\rm\em Proof of Tightness}
We denote $\bar{f}(q,d) = p_r$ in this proof for simplicity. 
The $d^*$ below is the optimal adversarial document defined in the proof of Lemma~\ref{lem:4.3}.
Note that $o_d = |T_d - T_{d^*}| / |T_d|$ and $1 - o_d = |T_d \cap T_{d^*}| / |T_d|$ as defined in Theorem~\ref{thm:4.1}.
Our proof is based on constructing a randomized smoothing ranker that satisfies the desired property we want to prove.

\header{Case 1 $p_r \leq 1 - o_d$} 
Note that in this case $|T_d \cap T_{d^*}| = 1 - o_d \geq p_r$.
Therefore, we can choose set $U$ such that $U \subseteq T_d \cap T_{d^*}$ and $|U|/|T_d| = p_r$. We define the ranker:
$$
 f_*(q,\mathbf{R})=\left\{
                \begin{array}{ll}
                  1, &\text{if}~\mathbf{R} \in U \cup T_{d^*}\\
                  0, &\text{otherwise}
                \end{array}
              \right. 
$$
\header{Case 2 $p_r > 1 - o_d$} 
We choose set $U$ such that $U \subseteq T_d \cap T_{d^*}$ and $|U| / |T_d| = p_r$.
We define the ranker 
$$
 f_*(q,\mathbf{R})=\left\{
                \begin{array}{ll}
                  1, &\text{if}~\mathbf{R} \in U \cup (T_{d^*} - T_d)\\
                  0, &\text{otherwise}
                \end{array}
              \right. 
$$

\noindent%
It can easily be verified that for each case, the defined ranker satisfies all the conditions in Theorem~\ref{thm:4.2}. This  indicates the bound can be achieved by learning a gold ranker that can judge some specific documents as relevant and others as irrelevant for the query.

\section{Experimental Setup}

% In this section, we introduce our experimental settings.

\subsection{Datasets}
To evaluate the effectiveness of our proposed methods, we conduct  experiments on two representative web search benchmark datasets. 

\begin{itemize}[leftmargin=*,nosep]
    \item \textbf{MS MARCO Document Ranking dataset} \cite{nguyen2016ms} (MS-MARCO-Doc) is a large-scale benchmark dataset for web document retrieval, with about 3.21M web documents and 0.37M training queries. The average length of the document is about 1129.
    \item \textbf{MS MARCO Passage Ranking dataset} \cite{nguyen2016ms} (MS-MARCO-Pas) is a large-scale benchmark dataset for passage retrieval, with about 8.84M passages from web pages and 0.5M training queries. The average length of the passage is about 58. 
  \end{itemize}

\subsection{Baselines}

Our CertDR can certify the top-$K$ robustness of different ranking models. We first  compare the certified top-$K$ robustness among different ranking models (i.e, BM25 \cite{BM25}, Duet \cite{Duet} and BERT \cite{devlin-etal-2019-bert}) under CertDR. Then, since the defense methods for NRMs have not been well explored yet, we adopt the representative empirical defense, i.e., \textbf{Data Augmentation (DA)}, in text classification task \cite{jia2017adversarial,ribeiro2018semantically}, for NRMs as a baseline. 
Specifically, for each training document $d$, we augment the collection with $2$ new documents $\Tilde{d}$ by sampling $\Tilde{d}$ uniformly from $S_d$, then train on the normal hinge loss following \cite{jia2019certified}. 
We do not use adversarial training \cite{FGSM} here because it would require running an adversarial search procedure at each training step, which would be prohibitively slow.

\subsection{Evaluation Metrics}
We evaluate the robustness to all WSRAs of models. 
We propose a metric to directly measure the \textbf{Certified Robust Query (CRQ)} percentage, the percentage of test queries for which the model is certified robust at the query $q$ if all the documents out of top-$K$ are not attacked into the top-$K$. 
Evaluating this exactly involves enumerating  exponentially many perturbations, which is intractable (Section \ref{sec: Certifying Smoothed Ranking Models}).
Instead, we evaluate the $CRQ$ under randomized smoothing, i.e.,
$$CRQ = \frac{\sum_{q \in Q} \mathbb{I} \{ \Delta L_q > 0 \}}{|Q|}, $$
where $\Delta L_q$ is the criterion mentioned in Section \ref{sec: Certifying Smoothed Ranking Models}. 
The ranking model is more certified robust with a higher CRQ (\%).

To compare the defense ability of CertDR with empirical defense methods, we also leverage two metrics, i.e., success rate and conditional success rate.
\textbf{Success Rate (SR) \cite{wu2022prada}} evaluates the percentage of the after-attack documents that are ranked higher than original documents. 
%   $$
%     \text{SR} = \frac{1}{|Q|}\sum_{t=1}^{|Q|}\frac{1}{N_{q}}\sum_{i=1}^{N_{q}}\mathbb{I}\{Rank_{L}(d_i + p) < Rank_{L}(d_i)\},
%     $$
%     where $|Q|$ denotes the number of evaluated queries, $N_{q}$ the number of attacked documents with respect to each query, and $d_i$ the attacked document with respect to the query $q \in Q$.
The robustness of a ranking model is better with a lower SR (\%).

Inspired by CondAcc \cite{wang2021certified}, which enables the comparison certified and empirical defense, we introduce  \textbf{Conditional Success Rate (CondSR)}. 
CondSR  evaluates whether the rankings of the adversarial documents in an attacked ranked list indeed cannot be improved when its counterpart clean ranked list is certified robust:
\begin{equation*}
\begin{split}   
&\text{CondSR} = {}\\[-2mm]
&\frac{\sum_{q \in Q} \mathbb{I} \{ \Delta L_q > 0 \} \frac{1}{N_{q}}\sum_{i=1}^{N_{q}}\mathbb{I}\{Rank_{L}(d_i + p) < Rank_{L}(d_i)\}}{\sum_{q \in Q} \mathbb{I} \{ \Delta L_q > 0 \}}.
\end{split}    
\end{equation*}
While CRQ is evaluated on clean ranked lists to show certified robustness, CondSR is tested on attacked ranked lists to show the empirical robustness of models on these certified ranked lists.
The robustness of a ranking model is better with a lower CondSR (\%).

\subsection{Implementation Details}

We implement ranking models following previous work \cite{wu2021neural, dai2019deeper}.
For the MS-MARCO-Doc collection, we use the official top 100 (i.e., $N=100$) ranked documents retrieved by the QL model. For the MS-MARCO-Pas, initial retrieval is performed using the Anserini toolkit \cite{yang2018anserini} with the BM25 model to obtain the top 100 ranked passages.
We evaluate all ranking models on 200 queries (i.e., $|Q|=200$) randomly sampled from the dev set of two datasets following \cite{wu2022prada}.
% For BM25, as suggested by Anserini, we set $k_1$ to 3.8 and $b$ to 0.87 for MS-MARCO-Doc, and set $k_1$ to 0.6 and $b$ to 0.62 for MS-MARCO-Pas, respectively.
% For Duet, we set the filter size as 32 in both local and distributed model and the hidden size as 32 in the fully-connected layer on two dataset following \cite{wu2021neural}.
% We apply BERT$_{base}$ released by Google following \cite{dai2019deeper} and choose the fine-tuning learning rate from $\{5e-5, 3e-5, 2e-5\}$ on two datasets.

% \header{Synonym Sets} We construct the synonym set $S_w$ of word $w$ to be the set of words with $\geq 0.9$ cosine similarity in the GLOVE vector space.
% The word vector space is constructed by post-processing the pre-trained GLOVE vectors \cite{pennington2014glove} using the counter-fitted method \cite{counter-fitted} and the ``all-but-the-top'' method \cite{mu2018all}, following \citet{ye2020safer}.

% \header{Perturbation Sets} In the collection perturbation dictionary $T_C$, two words $w$ and $w'$ are connected synonymously if there exists a path between the two words in the corresponding synonym network.
% Let $B_w$ to be the set of words connected to $w$ synonymously.
% Then we define the perturbation set $T_w$ to consist of the top $J$ words in $B_w$ with the largest GLOVE cosine similarity if $|B_w| \geq J$, and set $T_w = B_w$ if $|B_w| < J$. 
% Here $J$ is a hyper-parameter to control the size of $T_w$, and is set as 100 in this paper to get the best smoothness and ranking performance of $\bar{f}$. 

For the Monte Carlo estimation of $\Delta L_q $, we use 1,000 random perturbed documents to accept $\Delta L_q > 0$ with probability of at least 0.95.
The corresponding estimation error is 0.086 and is considered during the estimation following \citet{ye2020safer}. 
Further implementation details and the code can be found online.\footnote{\url{https://github.com/ict-bigdatalab/CertDR/}}

\section{Experimental Results}
We evaluate our defense method to address the following research questions: 
\begin{enumerate*}[label=(\textbf{RQ\arabic*})]
\item What is the certified robustness among different ranking models via CertDR? 
\item How does the randomized smoothed ranker perform compared with the original ranker?
\item How does $K$ affect certified top-$K$ robustness? 
\item How does CertDR perform compared with empirical defense baselines? 
\end{enumerate*}

% \item How does the number of nearest neighbors $J$ affect certified top-$K$ robustness and ranking performance when we construct the perturbation set $T_w$?
% Comparison with empirical methods

\subsection{Certified Top-$K$ Robustness of Different Ranking Models}
\label{Different Ranking Models}

\begin{figure}[t]
 \centering
% next to each other 
% \if0
\begin{tabular}{c}
 \subfigure[]{
 \includegraphics[clip,trim=2mm 0mm 0mm 0mm,width=0.455\columnwidth]{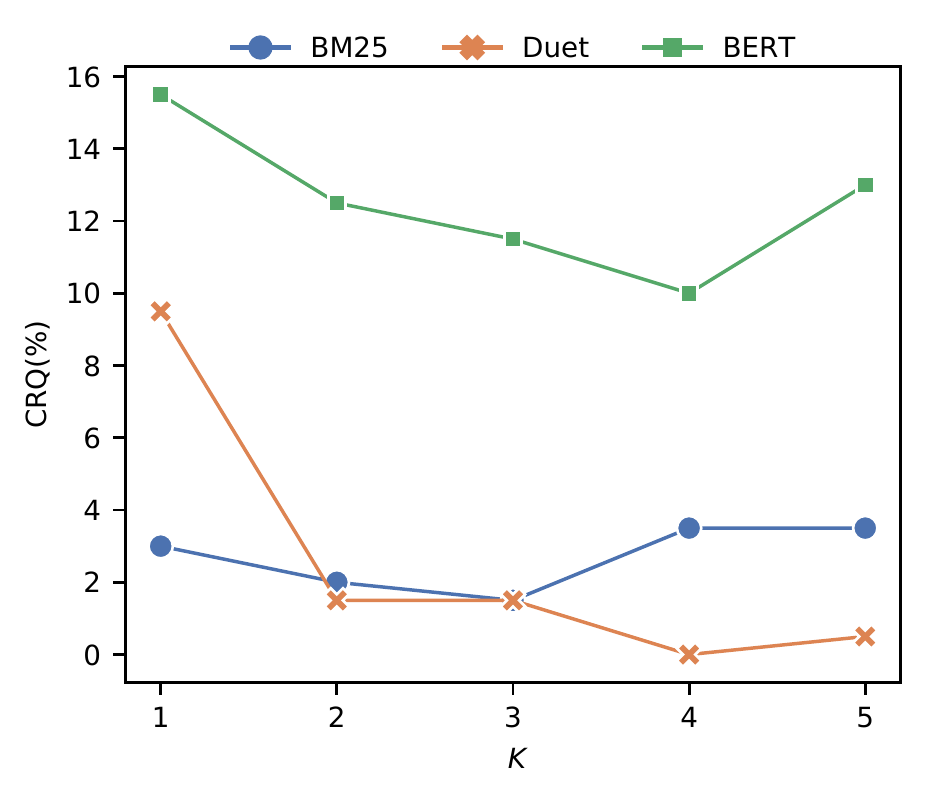}
 }
 \subfigure[]{
 \includegraphics[clip,trim=2mm 0mm 0mm 0mm,width=0.455\columnwidth]{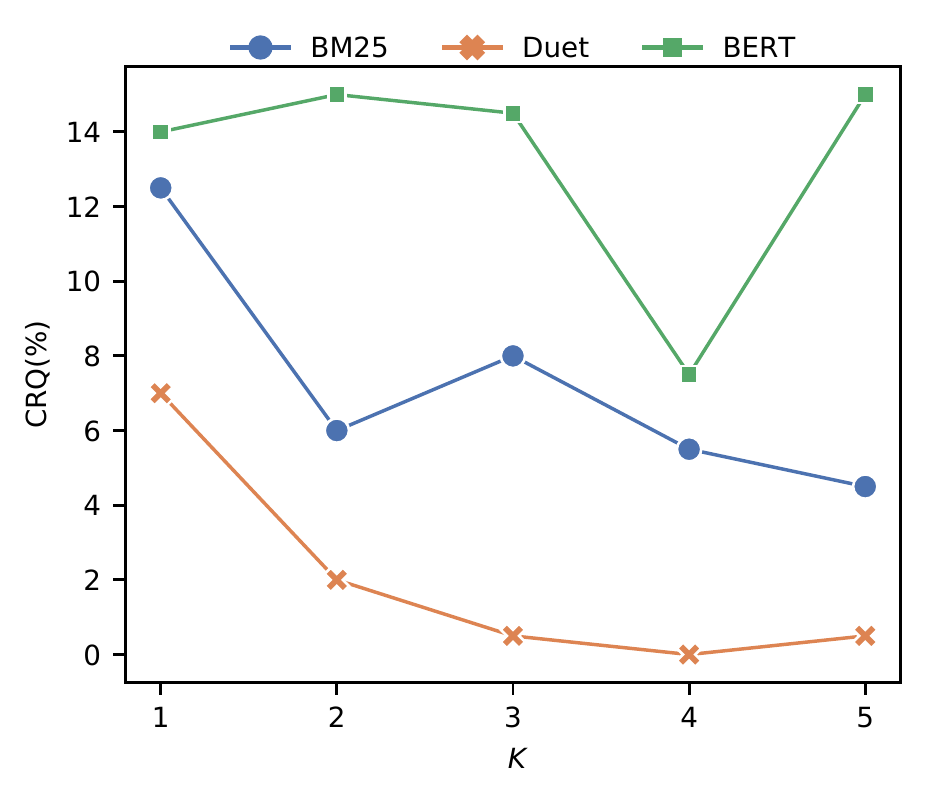}
 }
%  \fi
%below each other
\if0
 \includegraphics[clip,trim=0mm 0mm 0mm 0mm,width=0.6\columnwidth]{./figs/crq_marco_doc.pdf}
\\[-1ex]
 \includegraphics[clip,trim=0mm 0mm 0mm 0mm,width=0.6\columnwidth]{./figs/crq_marco_pas.pdf}
\\[-1ex]
\fi
\end{tabular}
 \caption{Certified top-$K$ robustness of different ranking models in terms of CRQ (\%) on MS-MARCO-Doc (a) and MS-MARCO-Pas (b). }
 \label{fig:crq_ranking_models}
\end{figure}

To answer \textbf{RQ1}, we analyze certified top-$K$ robustness of different ranking models using  CertDR on MS-MARCO-Doc and MS-MARCO-Pas.
See Figure \ref{fig:crq_ranking_models}. We find that
(i) Overall, the certified robustness of the ranking model is lower than that of the text classification models \cite{ye2020safer, jia2019certified}, indicating that ranking models are vulnerable to adversarial attacks. 
There are two potential reasons: 
First, the text ranking task itself needs to model cross-document interactions to capture query-document relevance, which is more complex than classifying a single sentence independently as in text classification. 
Second, certified top-$K$ robustness imposes requirements on the ranked list, which is demanding than the point-wise classification scenario.
%Therefore, it is challenging and worthwhile to explore the certified robustness for IR in future work. 
(ii) Pre-trained model BERT generally outperforms other models, indicating that BERT is more certified robust than other ranking models. 
The reason might be that pre-training on a large text corpus can improve the out-of-distribution generalizability to adversarial examples attacked by synonyms substitution. 
Therefore, it is worthwhile to leverage pre-training techniques to enhance the robustness of NRMs in the future. 
(iii) BM25 is less certified robust than Duet and BERT on MS-MARCO-Doc when $K$ is small, while it is more certified robust than Duet on MS-MARCO-Pas for all $K$s. 
BM25 depends on exact matching signals between the query and document; therefore, a possible explanation is that there are fewer options of attacked words in the short passage than the long document, contributing to the robustness on short texts. 
Besides, we leave the analysis of different $K$s to Section \ref{sec: analysis of k}. 

\begin{table}[h]
\centering
   \caption{Comparing the ranking performance between the original and  randomized smoothed ranker in terms of the MRR@10 and MRR@100 on MS-MARCO-Doc. * denotes significant degradation w.r.t.\ the randomized smoothed ranker w/o noise data augmentation (p-value<0.05). }
      \setlength\tabcolsep{7pt}
\renewcommand{\arraystretch}{1}
  	
  	\begin{tabular}{l c  c }
  \toprule
   Method & MRR@10 & MRR@100  \\
  \midrule
    Original $f$ & 0.4428$^{*}$ & 0.4470$^{*}$ \\
   Smoothed $\bar{f}$ w/o noise data aug & 0.2259\phantom{$^{*}$} & 0.2416\phantom{$^{*}$}\\
   Smoothed $\bar{f}$ & 0.3635$^{*}$ & 0.3722$^{*}$  \\
   \bottomrule
    \end{tabular}

   \label{table:ranking performance comparison}
\end{table}
 
 \vspace*{-1mm} 
\subsection{Smoothed Ranker vs.\ Original Ranker}

To answer \textbf{RQ2}, we compare the ranking performance of the randomized smoothed ranker with the original ranker.
We select BERT as the original ranker and conduct experiments on  MS-MARCO-Doc.
We also show the ranking performance of the randomized smoothed ranker without the noised data augmentation. 

From Table \ref{table:ranking performance comparison}, we observe that:
(i) The ranking performance of smoothed ranker without noised data augmentation drops dramatically (e.g., 0.2259 vs.\ 0.4428 in terms of MRR@100). 
The reason might be that the smoothed ranker ranks documents based on the ensemble ranking scores of perturbed documents, which are far away from the original documents. 
(ii) By applying a noised data augmentation strategy, the ranking performance of the smoothed ranker improves significantly and becomes closer to the original ranker. 
%This indicates that such strategy could train the ranker more effectively and correctly. 
The reason might be that the augmented training documents are generated from the same perturbation distribution with the perturbed documents, which helps the smoothed ranker learn to rank the perturbed documents properly. 
(iii) The smoothed ranker has a moderate drop in terms of MRR@100 compared with the original ranker with normal training (0.3722 vs. 0.4470). 
Similar drops on clean acc (accuracy on clean examples) are also seen for robust models in previous work \cite{miyato2016adversarial, jia2019certified}. 
Future work could explore how to achieve the trade-off between  clean and robust performance.

\begin{table}[h]
\centering
   \caption{The CRQ (\%) of different ranking models with different $K$ on MS-MARCO-Doc and MS-MARCO-Pas.}
      \setlength\tabcolsep{8pt}
%      \small
   \renewcommand{\arraystretch}{0.86}
  	\begin{tabular}{r c  c  c  c  c  c  c }
  \toprule
  \multirow{2}{*}{$K$} & \multicolumn{3}{c}{MS-MARCO-Doc} & \multicolumn{3}{c}{MS-MARCO-Pas} \\
   \cmidrule(r){2-4} \cmidrule(r){5-7}
  & BM25 & Duet & BERT & BM25 & Duet & BERT \\
   \midrule
  1 & 3.0 & 9.5 & 15.5 & 12.5 & 7.0 & 14.0 \\
  2 & 2.0 & 1.5 & 12.5 & 6.0 & 2.0 & 15.0 \\
  3 & 1.5 & 1.5 & 11.5 & 8.0 & 0.5 & 14.5 \\
  4 & 3.5 & 0 & 10.0 & 5.5 & 0 & 7.5 \\
  5 & 3.5 & 0.5 & 13.0 & 4.5 & 0.5 & 15.0 \\
  10 & 1.5 & 0 & 3.0 & 2.5 & 0 & 9.5 \\
  20 & 1.5 & 0 & 0 & 0.5 & 0 & 3.0 \\
  30 & 0.5 & 0 & 0 & 0 & 0 & 1.5 \\
  40 & 0.5 & 0 & 0 & 1.5 & 0 & 0 \\
  50 & 0.5 & 0 & 0 & 0.5 & 0 & 0 \\
  60 & 0.5 & 0 & 0 & 0 & 0 & 0 \\
  70 & 0.5 & 0 & 0 & 1.0 & 0 & 0 \\
  80 & 0.5 & 0 & 0 & 0 & 0 & 0 \\
  90 & 0.5 & 0 & 0 & 0.5 & 0 & 0 \\
\bottomrule
    \end{tabular}
    
   \label{table:analyzing different K}
   \vspace*{1mm}
\end{table}

\subsection{Analysis of the Effect of  $K$}

\label{sec: analysis of k}
% 需要解释一下为什么K大的甚至还有CQR高的
% 1、overall， K 大CQR变低，因为更容易攻破后面的
% 2、K大了CRQ变高的，比如xx和xx， 是因为第5篇更容易突破第4篇，所以xxx 可以直接拿marco pas 的4和5举例

To answer \textbf{RQ3}, we analyze the effect of $K$ for CertDR when we certify the top-$K$ robustness. 
Specifically, we analyze the ranking performance of different ranking models  in terms of CRQ, and set $K$ to 14 different values.
As shown in Table \ref{table:analyzing different K}, we can find that:
(i) Overall, the model becomes less certified top-$K$ robust  with the increase of $K$ on both datasets. 
Intuitively, it is more difficult to attack a document to a higher rank position than a lower rank position. 
(ii) However, an interesting finding is that the certified top-$K$ robustness with a larger $K$ is even greater than a smaller $K$ in a certain range.
For example, the CRQ of BERT is 15.0 with $K=5$ while 7.5 with $K=4$ on the MS-MARCO-Pas dataset. 
By conducting further analysis, we find that although documents ranked out of top $5$ are not easily attacked into the top $5$, the $5$-th document could be attacked into the top $4$ easily. 
(iii) While the CRQ of NRMs reduces to 0 after a point (e.g, the CRQ of Duet reduces to 0 after $K$=10), it is interesting to find that the CRQ of BM25 remains at a low positive value when $K$ is very large (e.g., CRQ of BM25 remains at 0.5 when $K$=30 to 90 on MS-MARCO-Doc). 
The reason may be that BM25 relies on the statistical features, which is more robust than word embeddings of NRMs against adversarial attacks. This is consistent with the findings in \citep{wu2021neural}.

% \subsection{Analysis of the Number of nearest neighbors $J$}
% \begin{table}[t]
% \centering
%   \caption{Needs to be added. BERT model }
%   \setlength\tabcolsep{2.1mm}

%   	\begin{tabular}{l | c  c  c  c  c  c  c c}
%   \toprule
%   J & 20 & 50 & 100 & 250 & 1000  \\
%   \midrule
%   CRQ (MS-MARCO-Doc) & 0 & 0 & 0 & 0 & 0  \\
%   CRQ (MS-MARCO-Pas) & 0 & 0 & 0 & 0 & 0  \\
% \bottomrule
%     \end{tabular}

%   \label{table:analyzing different J}
% \end{table}

\subsection{Comparison with Empirical Defenses}
Based on the above analysis of certified robustness of different models, we further compare CertDR with baseline empirical defense methods (i.e., DA) following \cite{wang2021certified}.
The WSRA is conducted by PRADA \cite{wu2022prada}, and we set $K=10$ and $5$ for CertDR on MS-MARCO-Doc and MS-MARCO-Pas, respectively. 

To answer \textbf{RQ4}, as shown in Table \ref{table:Baseline}, we observe that:
(i) The SR is very high (i.e., up to 96.7\% on MS-MARCO-Doc) if we do not take any defense, indicating that it is important to develop defense methods for NRMs to fight against adversarial attacks. 
(ii) Empirical defense method DA reduces the SR to some extent. 
However, it performs worse than CertDR.
Hence, simply augmenting the training documents (as in NLP) is not a robust defense against attacks in IR.
Future work should explore more adequate  empirical defense methods in IR.
Importantly, empirical defense methods do not provide rigorous certified robustness guarantees and the performance may significantly depend on the datasets and specific attacks.
(iii) CertDR achieves the lowest CondSR on both datasets, indicating that 
our CertDR could certify the robustness theoretically while enhancing the robustness empirically for ranking models.

\begin{table}[t]
\centering
   \caption{Comparisons between our proposed CertDR and the baseline on the BERT. Adversarial attacks are conducted by PRADA \cite{wu2022prada}. ADV corresponds to no defense. ADV and DA are evaluated under SR (\%) and CertDR is evaluated under CondSR (\%).}
   \setlength\tabcolsep{4.5mm}
      \renewcommand{\arraystretch}{1}
  	\begin{tabular}{l c c c }
  \toprule
  Dataset & ADV & DA & CertDR \\
  \midrule
  MS-MARCO-Doc & 96.7 & 57.0 & 40.0  \\
  MS-MARCO-Pas & 91.4 & 64.6 & 57.4  \\
\bottomrule
    \end{tabular}
   \label{table:Baseline}
   \vspace*{1mm}
\end{table}

\section{Conclusion}

In this paper, we defined the notion of Certified Top-$K$ Robustness for ranking models focusing on the characteristics of IR.
We proposed a certifiably robust defense method called CertDR, based on  randomized smoothing. 
The key idea is to smooth the ranking model with random word substitutions, and construct provable certification bounds based on the ranking property. 
Extensive experiments validate that CertDR outperforms existing defense methods and improves the certifiable robustness of ranking models. 

In future work, it is worth to strengthen the notion of Certified Top-$K$ Robustness to guarantee that the order of top-$K$ ranking results remains unchanged. 
We hope that our study helps to put concerns about the robustness of NRMs on the research agenda and to motivate new defense ideas, including empirical and certified defenses of ranking models.

\begin{acks}
This work was funded by the National Natural Science Foundation of China (NSFC) under Grants No. 62006218 and 61902381, the Youth Innovation Promotion Association CAS under Grants No. 20144310 and 2021100, the Innovation Project of ICT, CAS under Grants No. E261090, the Young Elite Scientist Sponsorship Program by CAST under Grants No. YESS20200121, and the Lenovo-CAS Joint Lab Youth Scientist Project.
This work was also (partially) funded by the Hybrid Intelligence Center, a 10-year program funded by the Dutch Ministry of Education, Culture and Science through the Dutch Research Council, \url{https://hybrid-intelligence-centre.nl}. All content represents the opinion of the authors, which is not necessarily shared or endorsed by their respective employers and/or sponsors.
\end{acks}

%\clearpage
\bibliographystyle{arxiv}
\balance
\bibliography{arxiv}

%%% -*-BibTeX-*-
%%% Do NOT edit. File created by BibTeX with style
%%% ACM-Reference-Format-Journals [18-Jan-2012].

\begin{thebibliography}{55}

%%% ====================================================================
%%% NOTE TO THE USER: you can override these defaults by providing
%%% customized versions of any of these macros before the \bibliography
%%% command.  Each of them MUST provide its own final punctuation,
%%% except for \shownote{}, \showDOI{}, and \showURL{}.  The latter two
%%% do not use final punctuation, in order to avoid confusing it with
%%% the Web address.
%%%
%%% To suppress output of a particular field, define its macro to expand
%%% to an empty string, or better, \unskip, like this:
%%%
%%% \newcommand{\showDOI}[1]{\unskip}   % LaTeX syntax
%%%
%%% \def \showDOI #1{\unskip}           % plain TeX syntax
%%%
%%% ====================================================================

\ifx \showCODEN    \undefined \def \showCODEN     #1{\unskip}     \fi
\ifx \showDOI      \undefined \def \showDOI       #1{#1}\fi
\ifx \showISBNx    \undefined \def \showISBNx     #1{\unskip}     \fi
\ifx \showISBNxiii \undefined \def \showISBNxiii  #1{\unskip}     \fi
\ifx \showISSN     \undefined \def \showISSN      #1{\unskip}     \fi
\ifx \showLCCN     \undefined \def \showLCCN      #1{\unskip}     \fi
\ifx \shownote     \undefined \def \shownote      #1{#1}          \fi
\ifx \showarticletitle \undefined \def \showarticletitle #1{#1}   \fi
\ifx \showURL      \undefined \def \showURL       {\relax}        \fi
% The following commands are used for tagged output and should be
% invisible to TeX
\providecommand\bibfield[2]{#2}
\providecommand\bibinfo[2]{#2}
\providecommand\natexlab[1]{#1}
\providecommand\showeprint[2][]{arXiv:#2}

\bibitem[\protect\citeauthoryear{Alzantot, Sharma, Elgohary, Ho, Srivastava,
  and Chang}{Alzantot et~al\mbox{.}}{2018}]%
        {alzantot2018generating}
\bibfield{author}{\bibinfo{person}{Moustafa Alzantot}, \bibinfo{person}{Yash
  Sharma}, \bibinfo{person}{Ahmed Elgohary}, \bibinfo{person}{Bo-Jhang Ho},
  \bibinfo{person}{Mani Srivastava}, {and} \bibinfo{person}{Kai-Wei Chang}.}
  \bibinfo{year}{2018}\natexlab{}.
\newblock \showarticletitle{Generating natural language adversarial examples}.
\newblock \bibinfo{journal}{\emph{arXiv preprint arXiv:1804.07998}}
  (\bibinfo{year}{2018}).
\newblock


\bibitem[\protect\citeauthoryear{Burges, Shaked, Renshaw, Lazier, Deeds,
  Hamilton, and Hullender}{Burges et~al\mbox{.}}{2005}]%
        {nDCG}
\bibfield{author}{\bibinfo{person}{Chris Burges}, \bibinfo{person}{Tal Shaked},
  \bibinfo{person}{Erin Renshaw}, \bibinfo{person}{Ari Lazier},
  \bibinfo{person}{Matt Deeds}, \bibinfo{person}{Nicole Hamilton}, {and}
  \bibinfo{person}{Greg Hullender}.} \bibinfo{year}{2005}\natexlab{}.
\newblock \showarticletitle{Learning to rank using gradient descent}. In
  \bibinfo{booktitle}{\emph{Proceedings of the 22nd international conference on
  Machine learning}}. \bibinfo{pages}{89--96}.
\newblock


\bibitem[\protect\citeauthoryear{Castillo and Davison}{Castillo and
  Davison}{2011}]%
        {castillo2011adversarial}
\bibfield{author}{\bibinfo{person}{Carlos Castillo} {and}
  \bibinfo{person}{Brian~D Davison}.} \bibinfo{year}{2011}\natexlab{}.
\newblock \bibinfo{booktitle}{\emph{Adversarial web search}}.
  Vol.~\bibinfo{volume}{4}.
\newblock


\bibitem[\protect\citeauthoryear{Cheng, Wei, and Hsieh}{Cheng
  et~al\mbox{.}}{2019}]%
        {cheng2019evaluating}
\bibfield{author}{\bibinfo{person}{Minhao Cheng}, \bibinfo{person}{Wei Wei},
  {and} \bibinfo{person}{Cho-Jui Hsieh}.} \bibinfo{year}{2019}\natexlab{}.
\newblock \showarticletitle{Evaluating and enhancing the robustness of dialogue
  systems: A case study on a negotiation agent}. In
  \bibinfo{booktitle}{\emph{NAACL}}.
\newblock


\bibitem[\protect\citeauthoryear{Cohen, Rosenfeld, and Kolter}{Cohen
  et~al\mbox{.}}{2019}]%
        {cohen2019certified}
\bibfield{author}{\bibinfo{person}{Jeremy Cohen}, \bibinfo{person}{Elan
  Rosenfeld}, {and} \bibinfo{person}{Zico Kolter}.}
  \bibinfo{year}{2019}\natexlab{}.
\newblock \showarticletitle{Certified adversarial robustness via randomized
  smoothing}. In \bibinfo{booktitle}{\emph{ICML}}.
\newblock


\bibitem[\protect\citeauthoryear{Dai and Callan}{Dai and Callan}{2019}]%
        {dai2019deeper}
\bibfield{author}{\bibinfo{person}{Zhuyun Dai} {and} \bibinfo{person}{Jamie
  Callan}.} \bibinfo{year}{2019}\natexlab{}.
\newblock \showarticletitle{Deeper text understanding for IR with contextual
  neural language modeling}. In \bibinfo{booktitle}{\emph{SIGIR}}.
  \bibinfo{pages}{985--988}.
\newblock


\bibitem[\protect\citeauthoryear{Devlin, Chang, Lee, and Toutanova}{Devlin
  et~al\mbox{.}}{2019}]%
        {devlin-etal-2019-bert}
\bibfield{author}{\bibinfo{person}{Jacob Devlin}, \bibinfo{person}{Ming-Wei
  Chang}, \bibinfo{person}{Kenton Lee}, {and} \bibinfo{person}{Kristina
  Toutanova}.} \bibinfo{year}{2019}\natexlab{}.
\newblock \showarticletitle{{BERT}: Pre-training of Deep Bidirectional
  Transformers for Language Understanding}. In
  \bibinfo{booktitle}{\emph{NAACL}}.
\newblock


\bibitem[\protect\citeauthoryear{Dong, Luu, Ji, and Liu}{Dong
  et~al\mbox{.}}{2020}]%
        {dong2020towards}
\bibfield{author}{\bibinfo{person}{Xinshuai Dong}, \bibinfo{person}{Anh~Tuan
  Luu}, \bibinfo{person}{Rongrong Ji}, {and} \bibinfo{person}{Hong Liu}.}
  \bibinfo{year}{2020}\natexlab{}.
\newblock \showarticletitle{Towards Robustness Against Natural Language Word
  Substitutions}. In \bibinfo{booktitle}{\emph{International Conference on
  Learning Representations}}.
\newblock


\bibitem[\protect\citeauthoryear{Dvijotham, Gowal, Stanforth, Arandjelovic,
  O'Donoghue, Uesato, and Kohli}{Dvijotham et~al\mbox{.}}{2018}]%
        {IBP}
\bibfield{author}{\bibinfo{person}{Krishnamurthy Dvijotham},
  \bibinfo{person}{Sven Gowal}, \bibinfo{person}{Robert Stanforth},
  \bibinfo{person}{Relja Arandjelovic}, \bibinfo{person}{Brendan O'Donoghue},
  \bibinfo{person}{Jonathan Uesato}, {and} \bibinfo{person}{Pushmeet Kohli}.}
  \bibinfo{year}{2018}\natexlab{}.
\newblock \showarticletitle{Training verified learners with learned verifiers}.
\newblock \bibinfo{journal}{\emph{arXiv preprint arXiv:1805.10265}}
  (\bibinfo{year}{2018}).
\newblock


\bibitem[\protect\citeauthoryear{Gao, Lanchantin, Soffa, and Qi}{Gao
  et~al\mbox{.}}{2018}]%
        {gao2018black}
\bibfield{author}{\bibinfo{person}{Ji Gao}, \bibinfo{person}{Jack Lanchantin},
  \bibinfo{person}{Mary~Lou Soffa}, {and} \bibinfo{person}{Yanjun Qi}.}
  \bibinfo{year}{2018}\natexlab{}.
\newblock \showarticletitle{Black-box generation of adversarial text sequences
  to evade deep learning classifiers}. In \bibinfo{booktitle}{\emph{SPW}}.
  IEEE, \bibinfo{pages}{50--56}.
\newblock


\bibitem[\protect\citeauthoryear{Goodfellow, Shlens, and Szegedy}{Goodfellow
  et~al\mbox{.}}{2014}]%
        {FGSM}
\bibfield{author}{\bibinfo{person}{Ian~J Goodfellow}, \bibinfo{person}{Jonathon
  Shlens}, {and} \bibinfo{person}{Christian Szegedy}.}
  \bibinfo{year}{2014}\natexlab{}.
\newblock \showarticletitle{Explaining and harnessing adversarial examples}.
\newblock \bibinfo{journal}{\emph{arXiv preprint arXiv:1412.6572}}
  (\bibinfo{year}{2014}).
\newblock


\bibitem[\protect\citeauthoryear{Goren, Kurland, Tennenholtz, and Raiber}{Goren
  et~al\mbox{.}}{2018}]%
        {goren2018ranking}
\bibfield{author}{\bibinfo{person}{Gregory Goren}, \bibinfo{person}{Oren
  Kurland}, \bibinfo{person}{Moshe Tennenholtz}, {and} \bibinfo{person}{Fiana
  Raiber}.} \bibinfo{year}{2018}\natexlab{}.
\newblock \showarticletitle{Ranking robustness under adversarial document
  manipulations}. In \bibinfo{booktitle}{\emph{SIGIR}}.
\newblock


\bibitem[\protect\citeauthoryear{Goren, Kurland, Tennenholtz, and Raiber}{Goren
  et~al\mbox{.}}{2020}]%
        {goren2020ranking}
\bibfield{author}{\bibinfo{person}{Gregory Goren}, \bibinfo{person}{Oren
  Kurland}, \bibinfo{person}{Moshe Tennenholtz}, {and} \bibinfo{person}{Fiana
  Raiber}.} \bibinfo{year}{2020}\natexlab{}.
\newblock \showarticletitle{Ranking-Incentivized Quality Preserving Content
  Modification}. In \bibinfo{booktitle}{\emph{SIGIR}}.
\newblock


\bibitem[\protect\citeauthoryear{Gu, Li, Liu, Ling, Su, Wei, and Zhu}{Gu
  et~al\mbox{.}}{2020}]%
        {gu2020speaker}
\bibfield{author}{\bibinfo{person}{Jia-Chen Gu}, \bibinfo{person}{Tianda Li},
  \bibinfo{person}{Quan Liu}, \bibinfo{person}{Zhen-Hua Ling},
  \bibinfo{person}{Zhiming Su}, \bibinfo{person}{Si Wei}, {and}
  \bibinfo{person}{Xiaodan Zhu}.} \bibinfo{year}{2020}\natexlab{}.
\newblock \showarticletitle{Speaker-aware BERT for multi-turn response
  selection in retrieval-based chatbots}. In \bibinfo{booktitle}{\emph{CIKM}}.
  \bibinfo{pages}{2041--2044}.
\newblock


\bibitem[\protect\citeauthoryear{Gyongyi and Garcia-Molina}{Gyongyi and
  Garcia-Molina}{2005}]%
        {gyongyi2005web}
\bibfield{author}{\bibinfo{person}{Zoltan Gyongyi} {and}
  \bibinfo{person}{Hector Garcia-Molina}.} \bibinfo{year}{2005}\natexlab{}.
\newblock \showarticletitle{Web spam taxonomy}. In
  \bibinfo{booktitle}{\emph{AIRWeb}}.
\newblock


\bibitem[\protect\citeauthoryear{Huang, Stanforth, Welbl, Dyer, Yogatama,
  Gowal, Dvijotham, and Kohli}{Huang et~al\mbox{.}}{2019}]%
        {huang2019achieving}
\bibfield{author}{\bibinfo{person}{Po-Sen Huang}, \bibinfo{person}{Robert
  Stanforth}, \bibinfo{person}{Johannes Welbl}, \bibinfo{person}{Chris Dyer},
  \bibinfo{person}{Dani Yogatama}, \bibinfo{person}{Sven Gowal},
  \bibinfo{person}{Krishnamurthy Dvijotham}, {and} \bibinfo{person}{Pushmeet
  Kohli}.} \bibinfo{year}{2019}\natexlab{}.
\newblock \showarticletitle{Achieving Verified Robustness to Symbol
  Substitutions via Interval Bound Propagation}. In
  \bibinfo{booktitle}{\emph{EMNLP/IJCNLP (1)}}.
\newblock


\bibitem[\protect\citeauthoryear{Jia and Liang}{Jia and Liang}{2017}]%
        {jia2017adversarial}
\bibfield{author}{\bibinfo{person}{Robin Jia} {and} \bibinfo{person}{Percy
  Liang}.} \bibinfo{year}{2017}\natexlab{}.
\newblock \showarticletitle{Adversarial Examples for Evaluating Reading
  Comprehension Systems}. In \bibinfo{booktitle}{\emph{EMNLP}}.
\newblock


\bibitem[\protect\citeauthoryear{Jia, Raghunathan, G{\"o}ksel, and Liang}{Jia
  et~al\mbox{.}}{2019}]%
        {jia2019certified}
\bibfield{author}{\bibinfo{person}{Robin Jia}, \bibinfo{person}{Aditi
  Raghunathan}, \bibinfo{person}{Kerem G{\"o}ksel}, {and}
  \bibinfo{person}{Percy Liang}.} \bibinfo{year}{2019}\natexlab{}.
\newblock \showarticletitle{Certified Robustness to Adversarial Word
  Substitutions.}. In \bibinfo{booktitle}{\emph{EMNLP/IJCNLP (1)}}.
\newblock


\bibitem[\protect\citeauthoryear{Joachims, Granka, Pan, Hembrooke, and
  Gay}{Joachims et~al\mbox{.}}{2017}]%
        {joachims2017accurately}
\bibfield{author}{\bibinfo{person}{Thorsten Joachims}, \bibinfo{person}{Laura
  Granka}, \bibinfo{person}{Bing Pan}, \bibinfo{person}{Helene Hembrooke},
  {and} \bibinfo{person}{Geri Gay}.} \bibinfo{year}{2017}\natexlab{}.
\newblock \showarticletitle{Accurately interpreting clickthrough data as
  implicit feedback}. In \bibinfo{booktitle}{\emph{Acm Sigir Forum}},
  Vol.~\bibinfo{volume}{51}. Acm New York, NY, USA, \bibinfo{pages}{4--11}.
\newblock


\bibitem[\protect\citeauthoryear{Joshi, Chen, Liu, Weld, Zettlemoyer, and
  Levy}{Joshi et~al\mbox{.}}{2020}]%
        {joshi2020spanbert}
\bibfield{author}{\bibinfo{person}{Mandar Joshi}, \bibinfo{person}{Danqi Chen},
  \bibinfo{person}{Yinhan Liu}, \bibinfo{person}{Daniel~S Weld},
  \bibinfo{person}{Luke Zettlemoyer}, {and} \bibinfo{person}{Omer Levy}.}
  \bibinfo{year}{2020}\natexlab{}.
\newblock \showarticletitle{SpanBERT: Improving pre-training by representing
  and predicting spans}.
\newblock \bibinfo{journal}{\emph{TACL}}  \bibinfo{volume}{8}
  (\bibinfo{year}{2020}), \bibinfo{pages}{64--77}.
\newblock


\bibitem[\protect\citeauthoryear{Katz, Barrett, Dill, Julian, and
  Kochenderfer}{Katz et~al\mbox{.}}{2017}]%
        {katz2017reluplex}
\bibfield{author}{\bibinfo{person}{Guy Katz}, \bibinfo{person}{Clark Barrett},
  \bibinfo{person}{David~L Dill}, \bibinfo{person}{Kyle Julian}, {and}
  \bibinfo{person}{Mykel~J Kochenderfer}.} \bibinfo{year}{2017}\natexlab{}.
\newblock \showarticletitle{Reluplex: An efficient SMT solver for verifying
  deep neural networks}. In \bibinfo{booktitle}{\emph{International conference
  on computer aided verification}}. Springer, \bibinfo{pages}{97--117}.
\newblock


\bibitem[\protect\citeauthoryear{Khattab and Zaharia}{Khattab and
  Zaharia}{2020}]%
        {ColBERT}
\bibfield{author}{\bibinfo{person}{Omar Khattab} {and} \bibinfo{person}{Matei
  Zaharia}.} \bibinfo{year}{2020}\natexlab{}.
\newblock \showarticletitle{Colbert: Efficient and effective passage search via
  contextualized late interaction over bert}. In
  \bibinfo{booktitle}{\emph{SIGIR}}.
\newblock


\bibitem[\protect\citeauthoryear{Lecuyer, Atlidakis, Geambasu, Hsu, and
  Jana}{Lecuyer et~al\mbox{.}}{2019}]%
        {lecuyer2019certified}
\bibfield{author}{\bibinfo{person}{Mathias Lecuyer}, \bibinfo{person}{Vaggelis
  Atlidakis}, \bibinfo{person}{Roxana Geambasu}, \bibinfo{person}{Daniel Hsu},
  {and} \bibinfo{person}{Suman Jana}.} \bibinfo{year}{2019}\natexlab{}.
\newblock \showarticletitle{Certified robustness to adversarial examples with
  differential privacy}. In \bibinfo{booktitle}{\emph{2019 IEEE Symposium on
  Security and Privacy (SP)}}. IEEE.
\newblock


\bibitem[\protect\citeauthoryear{Liang, Li, Su, Bian, Li, and Shi}{Liang
  et~al\mbox{.}}{2017}]%
        {liang2017deep}
\bibfield{author}{\bibinfo{person}{Bin Liang}, \bibinfo{person}{Hongcheng Li},
  \bibinfo{person}{Miaoqiang Su}, \bibinfo{person}{Pan Bian},
  \bibinfo{person}{Xirong Li}, {and} \bibinfo{person}{Wenchang Shi}.}
  \bibinfo{year}{2017}\natexlab{}.
\newblock \showarticletitle{Deep text classification can be fooled}.
\newblock \bibinfo{journal}{\emph{arXiv preprint arXiv:1704.08006}}
  (\bibinfo{year}{2017}).
\newblock


\bibitem[\protect\citeauthoryear{Lin, Nogueira, and Yates}{Lin
  et~al\mbox{.}}{2021}]%
        {lin2021pretrained}
\bibfield{author}{\bibinfo{person}{Jimmy Lin}, \bibinfo{person}{Rodrigo
  Nogueira}, {and} \bibinfo{person}{Andrew Yates}.}
  \bibinfo{year}{2021}\natexlab{}.
\newblock \showarticletitle{Pretrained transformers for text ranking: Bert and
  beyond}.
\newblock \bibinfo{journal}{\emph{Synthesis Lectures on Human Language
  Technologies}} \bibinfo{volume}{14}, \bibinfo{number}{4}
  (\bibinfo{year}{2021}), \bibinfo{pages}{1--325}.
\newblock


\bibitem[\protect\citeauthoryear{Liu}{Liu}{2011}]%
        {liu2011learning}
\bibfield{author}{\bibinfo{person}{Tie-Yan Liu}.}
  \bibinfo{year}{2011}\natexlab{}.
\newblock \bibinfo{booktitle}{\emph{Learning to Rank for Information
  Retrieval}}.
\newblock \bibinfo{publisher}{Springer Science \& Business Media}.
\newblock


\bibitem[\protect\citeauthoryear{Ma, Guo, Zhang, Fan, Li, and Cheng}{Ma
  et~al\mbox{.}}{2021}]%
        {ma2021b}
\bibfield{author}{\bibinfo{person}{Xinyu Ma}, \bibinfo{person}{Jiafeng Guo},
  \bibinfo{person}{Ruqing Zhang}, \bibinfo{person}{Yixing Fan},
  \bibinfo{person}{Yingyan Li}, {and} \bibinfo{person}{Xueqi Cheng}.}
  \bibinfo{year}{2021}\natexlab{}.
\newblock \showarticletitle{B-PROP: Bootstrapped pre-training with
  representative words prediction for ad-hoc retrieval}.
\newblock \bibinfo{journal}{\emph{arXiv preprint arXiv:2104.09791}}
  (\bibinfo{year}{2021}).
\newblock


\bibitem[\protect\citeauthoryear{Madry, Makelov, Schmidt, Tsipras, and
  Vladu}{Madry et~al\mbox{.}}{2017}]%
        {PGD}
\bibfield{author}{\bibinfo{person}{Aleksander Madry},
  \bibinfo{person}{Aleksandar Makelov}, \bibinfo{person}{Ludwig Schmidt},
  \bibinfo{person}{Dimitris Tsipras}, {and} \bibinfo{person}{Adrian Vladu}.}
  \bibinfo{year}{2017}\natexlab{}.
\newblock \showarticletitle{Towards deep learning models resistant to
  adversarial attacks}.
\newblock \bibinfo{journal}{\emph{arXiv preprint arXiv:1706.06083}}
  (\bibinfo{year}{2017}).
\newblock


\bibitem[\protect\citeauthoryear{Madry, Makelov, Schmidt, Tsipras, and
  Vladu}{Madry et~al\mbox{.}}{2018}]%
        {madry2018towards}
\bibfield{author}{\bibinfo{person}{Aleksander Madry},
  \bibinfo{person}{Aleksandar Makelov}, \bibinfo{person}{Ludwig Schmidt},
  \bibinfo{person}{Dimitris Tsipras}, {and} \bibinfo{person}{Adrian Vladu}.}
  \bibinfo{year}{2018}\natexlab{}.
\newblock \showarticletitle{Towards Deep Learning Models Resistant to
  Adversarial Attacks}. In \bibinfo{booktitle}{\emph{ICLR}}.
\newblock


\bibitem[\protect\citeauthoryear{Mitra, Diaz, and Craswell}{Mitra
  et~al\mbox{.}}{2017}]%
        {Duet}
\bibfield{author}{\bibinfo{person}{Bhaskar Mitra}, \bibinfo{person}{Fernando
  Diaz}, {and} \bibinfo{person}{Nick Craswell}.}
  \bibinfo{year}{2017}\natexlab{}.
\newblock \showarticletitle{Learning to match using local and distributed
  representations of text for web search}. In \bibinfo{booktitle}{\emph{WWW}}.
\newblock


\bibitem[\protect\citeauthoryear{Miyato, Dai, and Goodfellow}{Miyato
  et~al\mbox{.}}{2016}]%
        {miyato2016adversarial}
\bibfield{author}{\bibinfo{person}{Takeru Miyato}, \bibinfo{person}{Andrew~M
  Dai}, {and} \bibinfo{person}{Ian Goodfellow}.}
  \bibinfo{year}{2016}\natexlab{}.
\newblock \showarticletitle{Adversarial training methods for semi-supervised
  text classification}.
\newblock \bibinfo{journal}{\emph{arXiv preprint arXiv:1605.07725}}
  (\bibinfo{year}{2016}).
\newblock


\bibitem[\protect\citeauthoryear{Nguyen, Rosenberg, Song, Gao, Tiwary,
  Majumder, and Deng}{Nguyen et~al\mbox{.}}{2016}]%
        {nguyen2016ms}
\bibfield{author}{\bibinfo{person}{Tri Nguyen}, \bibinfo{person}{Mir
  Rosenberg}, \bibinfo{person}{Xia Song}, \bibinfo{person}{Jianfeng Gao},
  \bibinfo{person}{Saurabh Tiwary}, \bibinfo{person}{Rangan Majumder}, {and}
  \bibinfo{person}{Li Deng}.} \bibinfo{year}{2016}\natexlab{}.
\newblock \showarticletitle{MS MARCO: A human generated machine reading
  comprehension dataset}. In \bibinfo{booktitle}{\emph{CoCo@ NIPS}}.
\newblock


\bibitem[\protect\citeauthoryear{Niu, Guo, Lan, and Cheng}{Niu
  et~al\mbox{.}}{2012}]%
        {niu2012top}
\bibfield{author}{\bibinfo{person}{Shuzi Niu}, \bibinfo{person}{Jiafeng Guo},
  \bibinfo{person}{Yanyan Lan}, {and} \bibinfo{person}{Xueqi Cheng}.}
  \bibinfo{year}{2012}\natexlab{}.
\newblock \showarticletitle{Top-k learning to rank: labeling, ranking and
  evaluation}. In \bibinfo{booktitle}{\emph{SIGIR}}. \bibinfo{pages}{751--760}.
\newblock


\bibitem[\protect\citeauthoryear{Nogueira and Cho}{Nogueira and Cho}{2019}]%
        {monoBERT}
\bibfield{author}{\bibinfo{person}{Rodrigo Nogueira} {and}
  \bibinfo{person}{Kyunghyun Cho}.} \bibinfo{year}{2019}\natexlab{}.
\newblock \showarticletitle{Passage Re-ranking with BERT}.
\newblock \bibinfo{journal}{\emph{arXiv preprint arXiv:1901.04085}}
  (\bibinfo{year}{2019}).
\newblock


\bibitem[\protect\citeauthoryear{Ntoulas, Najork, Manasse, and
  Fetterly}{Ntoulas et~al\mbox{.}}{2006}]%
        {ntoulas2006detecting}
\bibfield{author}{\bibinfo{person}{Alexandros Ntoulas}, \bibinfo{person}{Marc
  Najork}, \bibinfo{person}{Mark Manasse}, {and} \bibinfo{person}{Dennis
  Fetterly}.} \bibinfo{year}{2006}\natexlab{}.
\newblock \showarticletitle{Detecting spam web pages through content analysis}.
  In \bibinfo{booktitle}{\emph{Proceedings of the 15th international conference
  on World Wide Web}}. \bibinfo{pages}{83--92}.
\newblock


\bibitem[\protect\citeauthoryear{Onal, Zhang, Altingovde, Rahman, Karagoz,
  Braylan, Dang, Chang, Kim, McNamara, Angert, Banner, Khetan, McDonnell,
  Nguyen, Xu, Wallace, de~Rijke, and Lease}{Onal et~al\mbox{.}}{2018}]%
        {onal-neural-2018}
\bibfield{author}{\bibinfo{person}{Kezban~Dilek Onal}, \bibinfo{person}{Ye
  Zhang}, \bibinfo{person}{Ismail~Sengor Altingovde},
  \bibinfo{person}{Md~Mustafizur Rahman}, \bibinfo{person}{Pinar Karagoz},
  \bibinfo{person}{Alex Braylan}, \bibinfo{person}{Brandon Dang},
  \bibinfo{person}{Heng-Lu Chang}, \bibinfo{person}{Henna Kim},
  \bibinfo{person}{Quinten McNamara}, \bibinfo{person}{Aaron Angert},
  \bibinfo{person}{Edward Banner}, \bibinfo{person}{Vivek Khetan},
  \bibinfo{person}{Tyler McDonnell}, \bibinfo{person}{An~Thanh Nguyen},
  \bibinfo{person}{Dan Xu}, \bibinfo{person}{Byron~C. Wallace},
  \bibinfo{person}{Maarten de Rijke}, {and} \bibinfo{person}{Matthew Lease}.}
  \bibinfo{year}{2018}\natexlab{}.
\newblock \showarticletitle{Neural information retrieval: At the end of the
  early years}.
\newblock \bibinfo{journal}{\emph{Information Retrieval Journal}}
  \bibinfo{volume}{21}, \bibinfo{number}{2--3} (\bibinfo{date}{June}
  \bibinfo{year}{2018}), \bibinfo{pages}{111--182}.
\newblock


\bibitem[\protect\citeauthoryear{Pennington, Socher, and Manning}{Pennington
  et~al\mbox{.}}{2014}]%
        {pennington2014glove}
\bibfield{author}{\bibinfo{person}{Jeffrey Pennington},
  \bibinfo{person}{Richard Socher}, {and} \bibinfo{person}{Christopher~D
  Manning}.} \bibinfo{year}{2014}\natexlab{}.
\newblock \showarticletitle{Glove: Global vectors for word representation}. In
  \bibinfo{booktitle}{\emph{EMNLP}}. \bibinfo{pages}{1532--1543}.
\newblock


\bibitem[\protect\citeauthoryear{Piskorski, Sydow, and Weiss}{Piskorski
  et~al\mbox{.}}{2008}]%
        {piskorski2008exploring}
\bibfield{author}{\bibinfo{person}{Jakub Piskorski}, \bibinfo{person}{Marcin
  Sydow}, {and} \bibinfo{person}{Dawid Weiss}.}
  \bibinfo{year}{2008}\natexlab{}.
\newblock \showarticletitle{Exploring linguistic features for web spam
  detection: a preliminary study}. In \bibinfo{booktitle}{\emph{Proceedings of
  the 4th international workshop on Adversarial information retrieval on the
  web}}. \bibinfo{pages}{25--28}.
\newblock


\bibitem[\protect\citeauthoryear{Ponte and Croft}{Ponte and Croft}{1998}]%
        {QL}
\bibfield{author}{\bibinfo{person}{Jay~M Ponte} {and} \bibinfo{person}{W~Bruce
  Croft}.} \bibinfo{year}{1998}\natexlab{}.
\newblock \showarticletitle{A language modeling approach to information
  retrieval}. In \bibinfo{booktitle}{\emph{SIGIR}}. \bibinfo{pages}{275--281}.
\newblock


\bibitem[\protect\citeauthoryear{Radev, Qi, Wu, and Fan}{Radev
  et~al\mbox{.}}{2002}]%
        {MRR}
\bibfield{author}{\bibinfo{person}{Dragomir~R Radev}, \bibinfo{person}{Hong
  Qi}, \bibinfo{person}{Harris Wu}, {and} \bibinfo{person}{Weiguo Fan}.}
  \bibinfo{year}{2002}\natexlab{}.
\newblock \showarticletitle{Evaluating Web-based Question Answering Systems.}.
  In \bibinfo{booktitle}{\emph{LREC}}. Citeseer.
\newblock


\bibitem[\protect\citeauthoryear{Ren, Deng, He, and Che}{Ren
  et~al\mbox{.}}{2019}]%
        {ren2019generating}
\bibfield{author}{\bibinfo{person}{Shuhuai Ren}, \bibinfo{person}{Yihe Deng},
  \bibinfo{person}{Kun He}, {and} \bibinfo{person}{Wanxiang Che}.}
  \bibinfo{year}{2019}\natexlab{}.
\newblock \showarticletitle{Generating natural language adversarial examples
  through probability weighted word saliency}. In
  \bibinfo{booktitle}{\emph{ACL}}. \bibinfo{pages}{1085--1097}.
\newblock


\bibitem[\protect\citeauthoryear{Ribeiro, Singh, and Guestrin}{Ribeiro
  et~al\mbox{.}}{2018}]%
        {ribeiro2018semantically}
\bibfield{author}{\bibinfo{person}{Marco~Tulio Ribeiro},
  \bibinfo{person}{Sameer Singh}, {and} \bibinfo{person}{Carlos Guestrin}.}
  \bibinfo{year}{2018}\natexlab{}.
\newblock \showarticletitle{Semantically equivalent adversarial rules for
  debugging NLP models}. In \bibinfo{booktitle}{\emph{ACL}}.
\newblock


\bibitem[\protect\citeauthoryear{Robertson and Walker}{Robertson and
  Walker}{1994}]%
        {BM25}
\bibfield{author}{\bibinfo{person}{Stephen~E Robertson} {and}
  \bibinfo{person}{Steve Walker}.} \bibinfo{year}{1994}\natexlab{}.
\newblock \showarticletitle{Some simple effective approximations to the
  2-poisson model for probabilistic weighted retrieval}. In
  \bibinfo{booktitle}{\emph{SIGIR’94}}. Springer, \bibinfo{pages}{232--241}.
\newblock


\bibitem[\protect\citeauthoryear{Singh, Ganvir, P{\"u}schel, and Vechev}{Singh
  et~al\mbox{.}}{2019}]%
        {singh2019beyond}
\bibfield{author}{\bibinfo{person}{Gagandeep Singh}, \bibinfo{person}{Rupanshu
  Ganvir}, \bibinfo{person}{Markus P{\"u}schel}, {and} \bibinfo{person}{Martin
  Vechev}.} \bibinfo{year}{2019}\natexlab{}.
\newblock \showarticletitle{Beyond the single neuron convex barrier for neural
  network certification}.
\newblock \bibinfo{journal}{\emph{NIPS}} (\bibinfo{year}{2019}).
\newblock


\bibitem[\protect\citeauthoryear{Szegedy, Zaremba, Sutskever, Bruna, Erhan,
  Goodfellow, and Fergus}{Szegedy et~al\mbox{.}}{2013}]%
        {szegedy2013intriguing}
\bibfield{author}{\bibinfo{person}{Christian Szegedy},
  \bibinfo{person}{Wojciech Zaremba}, \bibinfo{person}{Ilya Sutskever},
  \bibinfo{person}{Joan Bruna}, \bibinfo{person}{Dumitru Erhan},
  \bibinfo{person}{Ian Goodfellow}, {and} \bibinfo{person}{Rob Fergus}.}
  \bibinfo{year}{2013}\natexlab{}.
\newblock \showarticletitle{Intriguing properties of neural networks}.
\newblock \bibinfo{journal}{\emph{arXiv preprint arXiv:1312.6199}}
  (\bibinfo{year}{2013}).
\newblock


\bibitem[\protect\citeauthoryear{Wang, Tang, Lou, and Xiong}{Wang
  et~al\mbox{.}}{2021}]%
        {wang2021certified}
\bibfield{author}{\bibinfo{person}{Wenjie Wang}, \bibinfo{person}{Pengfei
  Tang}, \bibinfo{person}{Jian Lou}, {and} \bibinfo{person}{Li Xiong}.}
  \bibinfo{year}{2021}\natexlab{}.
\newblock \showarticletitle{Certified robustness to word substitution attack
  with differential privacy}. In \bibinfo{booktitle}{\emph{Proceedings of the
  2021 Conference of the North American Chapter of the Association for
  Computational Linguistics: Human Language Technologies}}.
  \bibinfo{pages}{1102--1112}.
\newblock


\bibitem[\protect\citeauthoryear{Wang, Ma, Bailey, Yi, Zhou, and Gu}{Wang
  et~al\mbox{.}}{2019}]%
        {wang2019convergence}
\bibfield{author}{\bibinfo{person}{Yisen Wang}, \bibinfo{person}{Xingjun Ma},
  \bibinfo{person}{James Bailey}, \bibinfo{person}{Jinfeng Yi},
  \bibinfo{person}{Bowen Zhou}, {and} \bibinfo{person}{Quanquan Gu}.}
  \bibinfo{year}{2019}\natexlab{}.
\newblock \showarticletitle{On the Convergence and Robustness of Adversarial
  Training}. In \bibinfo{booktitle}{\emph{ICML}}.
\newblock


\bibitem[\protect\citeauthoryear{Wu, Zhang, Guo, de~Rijke, Fan, and Cheng}{Wu
  et~al\mbox{.}}{2022}]%
        {wu2022prada}
\bibfield{author}{\bibinfo{person}{Chen Wu}, \bibinfo{person}{Ruqing Zhang},
  \bibinfo{person}{Jiafeng Guo}, \bibinfo{person}{Maarten de Rijke},
  \bibinfo{person}{Yixing Fan}, {and} \bibinfo{person}{Xueqi Cheng}.}
  \bibinfo{year}{2022}\natexlab{}.
\newblock \showarticletitle{PRADA: Practical Black-Box Adversarial Attacks
  against Neural Ranking Models}.
\newblock \bibinfo{journal}{\emph{arXiv preprint arXiv:2204.01321}}
  (\bibinfo{year}{2022}).
\newblock


\bibitem[\protect\citeauthoryear{Wu, Zhang, Guo, Fan, and Cheng}{Wu
  et~al\mbox{.}}{2021}]%
        {wu2021neural}
\bibfield{author}{\bibinfo{person}{Chen Wu}, \bibinfo{person}{Ruqing Zhang},
  \bibinfo{person}{Jiafeng Guo}, \bibinfo{person}{Yixing Fan}, {and}
  \bibinfo{person}{Xueqi Cheng}.} \bibinfo{year}{2021}\natexlab{}.
\newblock \showarticletitle{Are Neural Ranking Models Robust?}
\newblock \bibinfo{journal}{\emph{arXiv preprint arXiv:2108.05018}}
  (\bibinfo{year}{2021}).
\newblock


\bibitem[\protect\citeauthoryear{Xia, Liu, and Li}{Xia et~al\mbox{.}}{2009}]%
        {xia2009statistical}
\bibfield{author}{\bibinfo{person}{Fen Xia}, \bibinfo{person}{Tie-Yan Liu},
  {and} \bibinfo{person}{Hang Li}.} \bibinfo{year}{2009}\natexlab{}.
\newblock \showarticletitle{Statistical consistency of top-k ranking}.
\newblock \bibinfo{journal}{\emph{NIPS}}  \bibinfo{volume}{22}
  (\bibinfo{year}{2009}).
\newblock


\bibitem[\protect\citeauthoryear{Yang, Fang, and Lin}{Yang
  et~al\mbox{.}}{2018}]%
        {yang2018anserini}
\bibfield{author}{\bibinfo{person}{Peilin Yang}, \bibinfo{person}{Hui Fang},
  {and} \bibinfo{person}{Jimmy Lin}.} \bibinfo{year}{2018}\natexlab{}.
\newblock \showarticletitle{Anserini: Reproducible ranking baselines using
  Lucene}.
\newblock \bibinfo{journal}{\emph{Journal of Data and Information Quality
  (JDIQ)}} \bibinfo{volume}{10}, \bibinfo{number}{4} (\bibinfo{year}{2018}),
  \bibinfo{pages}{1--20}.
\newblock


\bibitem[\protect\citeauthoryear{Ye, Gong, and Liu}{Ye et~al\mbox{.}}{2020}]%
        {ye2020safer}
\bibfield{author}{\bibinfo{person}{Mao Ye}, \bibinfo{person}{Chengyue Gong},
  {and} \bibinfo{person}{Qiang Liu}.} \bibinfo{year}{2020}\natexlab{}.
\newblock \showarticletitle{SAFER: A structure-free approach for certified
  robustness to adversarial word substitutions}.
\newblock \bibinfo{journal}{\emph{arXiv preprint arXiv:2005.14424}}
  (\bibinfo{year}{2020}).
\newblock


\bibitem[\protect\citeauthoryear{Zang, Qi, Yang, Liu, Zhang, Liu, and Sun}{Zang
  et~al\mbox{.}}{2019}]%
        {zang2019word}
\bibfield{author}{\bibinfo{person}{Yuan Zang}, \bibinfo{person}{Fanchao Qi},
  \bibinfo{person}{Chenghao Yang}, \bibinfo{person}{Zhiyuan Liu},
  \bibinfo{person}{Meng Zhang}, \bibinfo{person}{Qun Liu}, {and}
  \bibinfo{person}{Maosong Sun}.} \bibinfo{year}{2019}\natexlab{}.
\newblock \showarticletitle{Word-level textual adversarial attacking as
  combinatorial optimization}.
\newblock \bibinfo{journal}{\emph{arXiv preprint arXiv:1910.12196}}
  (\bibinfo{year}{2019}).
\newblock


\bibitem[\protect\citeauthoryear{Zhang, Sheng, Alhazmi, and Li}{Zhang
  et~al\mbox{.}}{2020}]%
        {zhang2020adversarial}
\bibfield{author}{\bibinfo{person}{Wei~Emma Zhang}, \bibinfo{person}{Quan~Z
  Sheng}, \bibinfo{person}{Ahoud Alhazmi}, {and} \bibinfo{person}{Chenliang
  Li}.} \bibinfo{year}{2020}\natexlab{}.
\newblock \showarticletitle{Adversarial attacks on deep-learning models in
  natural language processing: A survey}.
\newblock \bibinfo{journal}{\emph{TIST}} \bibinfo{volume}{11},
  \bibinfo{number}{3} (\bibinfo{year}{2020}), \bibinfo{pages}{1--41}.
\newblock


\bibitem[\protect\citeauthoryear{Zhou, Niu, Wang, Zhang, and Hua}{Zhou
  et~al\mbox{.}}{2020}]%
        {zhou2020adversarial}
\bibfield{author}{\bibinfo{person}{Mo Zhou}, \bibinfo{person}{Zhenxing Niu},
  \bibinfo{person}{Le Wang}, \bibinfo{person}{Qilin Zhang}, {and}
  \bibinfo{person}{Gang Hua}.} \bibinfo{year}{2020}\natexlab{}.
\newblock \showarticletitle{Adversarial ranking attack and defense}. In
  \bibinfo{booktitle}{\emph{ECCV}}.
\newblock


\end{thebibliography}

\end{document}